\documentclass[a4paper,11pt]{article}
\pdfoutput=1 

\usepackage{jheppub} 

\usepackage[T1]{fontenc} 
\usepackage[english]{babel}
\usepackage{bbm}
\usepackage{physics}
\usepackage[usenames,dvipsnames]{color}
\usepackage[dvipsnames]{xcolor}

\usepackage{hyperref}
\hypersetup{
	pdftitle    = {Conformal field theory complexity from Euler-Arnold equations},
	pdfauthor   = {Mario Flory and Michal P.~Heller}
}

\renewcommand{\Re}{\text{Re}}
\renewcommand{\Im}{\text{Im}}

\newcommand{\mA}{\mathcal{A}}

\newcommand{\mL}{\mathcal{L}}

\newcommand{\mT}{\mathcal{T}}
\newcommand{\mR}{\mathcal{R}}

\newcommand{\mO}{\mathcal{O}}

\newcommand{\mQ}{\mathcal{Q}}

\usepackage[bordercolor=black,backgroundcolor=yellow,linecolor=black,textsize=scriptsize,shadow]{todonotes}

\newcommand{\hr}{\middle| h \right\rangle }
\newcommand{\hl}{\left\langle h \middle|}
\newcommand{\hrr}{\left| h \right\rangle }
\newcommand{\hll}{\left\langle h \right|}

\newcommand{\tcs}{\tau,\sigma}
\newcommand{\tck}{\tau,\kappa}

\newcommand{\qed}{\nobreak \ifvmode \relax \else\ifdim\lastskip<1.5em 
\hskip-\lastskip\hskip1.5em plus0em minus0.5em \fi \nobreak\vrule height0.75em 
width0.5em depth0.25em\fi}

\newcommand{\be}{\begin{eqnarray}}
\newcommand{\ee}{\end{eqnarray}}

\def\>{\rangle}
\def\<{\langle}

\newcommand{\executeiffilenewer}[3]{%
\ifnum\pdfstrcmp{\pdffilemoddate{#1}}%
{\pdffilemoddate{#2}}>0%
{\immediate\write18{#3}}\fi%
}

\graphicspath{{pics/}}

\usepackage[colorlinks = true,
            linkcolor = blue,
            urlcolor  = blue,
            citecolor = blue,
            anchorcolor = blue]{hyperref}

\title{\boldmath Conformal field theory complexity from Euler-Arnold equations}

\author[a]{Mario Flory}
\author[b]{and Michal P.~Heller}

\affiliation[a]{Institute of Physics, Jagiellonian University, \L{}ojasiewicza 11, 30-348 Krak\'ow, Poland}
\affiliation[b]{Max Planck Institute for Gravitational Physics (Albert Einstein Institute), Am M{\"u}hlenberg 1, 14476 Potsdam-Golm, Germany
\\
On leave from: \textit{National Centre for Nuclear Research, Pasteura 7, 02-093 Warsaw, Poland}}

\emailAdd{mflory@th.if.uj.edu.pl}
\emailAdd{michal.p.heller@aei.mpg.de}

\abstract{Defining complexity in quantum field theory is a difficult task, and the main challenge concerns going beyond free models and associated Gaussian states and operations. One take on this issue is to consider conformal field theories in 1+1 dimensions and our work is a comprehensive study of state and operator complexity in the universal sector of their energy-momentum tensor. The unifying conceptual ideas are Euler-Arnold equations and their integro-differential generalization, which guarantee well-posedness of the optimization problem between two generic states or transformations of interest. The present work provides an in-depth discussion of the results reported in~\href{https://arxiv.org/abs/2005.02415}{arXiv:2005.02415} and techniques used in their derivation. Among the most important topics we cover are usage of differential regularization, solution of the integro-differential equation describing Fubini-Study state complexity and probing the underlying geometry.}

\keywords{AdS-CFT Correspondence, Gauge-gravity correspondence, Conformal Field Theory}

\begin{document} 
\maketitle
\flushbottom

\setcounter{tocdepth}{2}
\thispagestyle{empty}


\section{Introduction}

\label{sec::Intro}

The interplay between quantum field theory (QFT) and geometry has a long and rich history. Recently, a new and promising instance has emerged, which concerns geometrization of state preparation or representations of, primarily, the time evolution operator in systems described by quantum fields. In this approach one seeks to find a way of preparing a state or an operator of interest minimizing the use of what one regards as more primitive operators or states by relating it to a geodesic problem in an auxiliary geometry. This introduces then a relation between different operators or states according to their \emph{complexity} -- the length of the corresponding minimal geodesic.

The approach elucidated above was originally introduced in \cite{Nielsen:2005mn1,Nielsen:2006mn2,Nielsen:2007mn3} as a way of bounding qubit circuits preparing unitary transformations of interest in the context of quantum computing and currently serves as our best definition of complexity in the QFT setting. In the course of the past three years starting with  \cite{Chapman:2017rqy,Jefferson:2017sdb} many results were delivered on complexity of quantum fields by focusing in the vast majority of cases on free QFTs soluble via Gaussian techniques~\cite{Weedbrook2012}, see \cite{Khan:2018rzm,Hackl:2018ptj,Camargo:2018eof,Chapman:2018hou,Guo:2018kzl,Liu:2019aji,Bernamonti:2019zyy,Caceres:2019pgf,Braccia:2019xxi,Bernamonti:2020bcf,DiGiulio:2020hlz,Guo:2020dsi} for sample developments.

To be more specific, the approach of~\cite{Chapman:2017rqy,Jefferson:2017sdb} in its default setting concerns unitary transformations $U$ represented as a sequence of steps
\begin{equation}
\label{eq.circuit}
U(\tau) =  \overleftarrow{{\cal P}} e^{- i \int_{0}^{\tau}  Q(\gamma)d \gamma} \quad \mathrm{with} \quad U(\tau = 1) \equiv U.
\end{equation}
A unitary $U$ can be either a standalone object of interest or a tool to be used to map some reference state~$|\mR\rangle$ to a desired target state~$|\mT\rangle$,
\begin{equation}
\label{eq.TUR}
|\mT\rangle = U\, |\mR\rangle,
\end{equation}
via a sequence of intermediate states
\begin{equation}
\label{psioftau}
|\psi(\tau)\rangle \equiv U(\tau) \, |\mR \rangle.
\end{equation}
The key idea introduced in~\cite{Nielsen:2005mn1,Nielsen:2006mn2,Nielsen:2007mn3} originating from the quantum optimal control theory~\cite{werschnik2007quantum} is to view Hermitian generators of circuit layers~$Q(\tau)$ as being by themselves obtained from more elementary Hermitian building blocks ${\cal O}_{I}$,
\begin{equation}
\label{eq.deflayergen}
Q(\tau) = \sum_{I} {\cal O}_{I} \, Y^{I} (\tau), 
\end{equation}
where real functions $Y^{I}(\tau)$ multiplied by $d\tau$ 
represent intuitively the number of times a layer between $\tau$ and $\tau + d\tau$ utilized the operation encapsulated by ${\cal O}_{I}$. Following this logic, it is then natural to associate $\int_{0}^{1} d\tau \sum_{I} |Y^{I}|$ with the cost of a given realization of~$U$ and define complexity $\cal C$ as the minimum of such costs subject to the condition $U(\tau = 1) = U$. In this definition we used the Manhattan norm, but other choices are possible, in particular various incarnations and generalizations of a Euclidean norm, to define complexity. One such option is based on evaluating the variance of the generator $Q(\tau)$ in some auxiliary state $|\mA\rangle$,
\begin{equation}
\label{eq.costvar}
\mathrm{L}_{\mathrm{var}} = \int_{0}^{1} d\tau \, \sqrt{ \langle \mA| Q(\tau)^2 | \mA \rangle - \langle \mA| Q(\tau) | \mA \rangle^2 },
\end{equation}
with complexity being the minimum of~$\mathrm{L}_{\mathrm{var}}$. One should note that all the above choices lead to in principle independent definitions of complexity.

The discussion so far concerned the complexity of operators, but, as in~\cite{Jefferson:2017sdb}, one can also use it to define a~complexity of states scanning over all $U$ represented as a circuit~\eqref{eq.circuit} subject to the condition~\eqref{eq.TUR}. There exists also a standalone definition of complexity of states, which is defined similarly to~\eqref{eq.costvar}, but evaluates the variance in the state $|\psi(\tau)\rangle$
\begin{equation}
\label{eq.costFS}
\mathrm{L}_{\mathrm{FS}} = \int_{0}^{1} d\tau \sqrt{ \langle \psi(\tau)| Q(\tau)^2 | \psi(\tau) \rangle - \langle \psi(\tau)| Q(\tau) | \psi(\tau) \rangle^2 }.
\end{equation}
Such a definition was introduced in~\cite{Chapman:2017rqy} as the Fubini-Study complexity, since it is effectively based on the distance traversed in the Hilbert space according to the Hilbert space metric, the Fubini-Study distance~\cite{geomquantstates}, by the circuit~\eqref{eq.circuit} acting on a reference state $|\mR\rangle$. The key difference between this notion and the ones based on unitary complexity is that in the Fubini-Study approach the cost associated with a given layer depends on where it is placed in the circuit, since the variance changes as the circuit progresses. In the present article, as in~\cite{Flory:2020eot}, we will take the Fubini-Study complexity as our point of departure and guided by it we will also arrive at several versions of complexity of unitary operators.

The initial motivation to study complexity in the QFT realm stems from another example of geometrization of QFT -- holography, also known as AdS/CFT or gauge-gravity duality~\cite{Maldacena:1997re,Witten:1998qj,Gubser:1998bc}. In holography, qualitative yet convincing considerations building primarily on an analogy between tensor networks~\cite{Orus:2014poa} describing states in underlying QFTs and dual geometries~\cite{Swingle:2009bg} have led to the interpretation of novel gravitational observables -- certain codimension-one and -zero volumes and gravitational action in causally defined regions -- as the aforementioned complexity~\cite{Susskind:2014rva,Stanford:2014jda,Couch:2016exn,Brown:2015bva,Brown:2015lvg}. While the subsequent studies of complexity in free QFTs have matched some of the predictions of these so-called holographic complexity proposals, they have failed to reproduce others for a good reason -- holographic QFTs are the complete opposite to being free.

As we alluded, the reason why free QFTs are simple stems from the fact that interesting states including ground states or thermofield double states are Gaussian and can be constructed from simpler Gaussian states in terms of the group of Gaussian transformations. In such settings, the field theory is typically represented on a lattice and operators ${\cal O}_{I}$ are considered to be all operators quadratic in latticized fields. Such Gaussian operators generate the symplectic group in the case of free bosons~\cite{Chapman:2018hou} and the special orthogonal group in the case of free fermions~\cite{Khan:2018rzm,Hackl:2018ptj}. Using one of the Euclidean-type cost functions assigns a non-negative definite right-invariant metric on such a Lie group and the problem of finding the optimal circuit reduces to finding the minimal geodesic in these geometries, as advertized above. 

One clear way to make progress within the framework of~\cite{Chapman:2017rqy,Jefferson:2017sdb} is to replace the group of Gaussian transformations by a group that is realized in holographic QFTs and is powerful enough to allow us to access interesting states\footnote{A similar but different line of research concerns the "first law of complexity" \cite{Bernamonti:2019zyy,Bernamonti:2020bcf,Hashemi:2019aop}, wherein (infinitesimal) complexity changes in bulk and boundary calculations can be compared if they are generated by acting on the state with a well understood generator.}. Such a group is given by the Virasoro group generating local conformal transformations across (1+1)-dimensional conformal field theories (CFT$_{1+1}$). This realization goes back to~\cite{Caputa:2017urj,Caputa:2017yrh,Czech:2017ryf,Caputa:2018kdj,Belin:2018bpg,Flory:2018akz,Flory:2019kah,Camargo:2019isp,Chen:2020nlj,Erdmenger:2020sup}, see in particular~\cite{Magan:2018nmu} for an early study of complexity associated with symmetry group transformations in a~QFT. It is also the topic of the present work, as well as our previous paper~\cite{Flory:2020eot},
where we introduce a complexity notion based on the Fubini-Study metric to the framework of~\cite{Caputa:2018kdj,Erdmenger:2020sup}. Given the agreement between the bulk result of \cite{Flory:2018akz} and the field theory result of \cite{Belin:2018bpg} for the leading order of complexity change under infinitesimal local transformations, we hope that a better understanding of Fubini-Study complexity on the Virasoro group might help to better understand and check the entry in the holographic dictionary relating bulk volumes and boundary complexity. 

The goal of this paper is to present our results in a manner that is self contained and accessible for readers who approach the topic from a holographic complexity point of view, while also making connections with relevant parts of the physical and mathematical literature that may not be widely known or appreciated in the complexity community. This applies especially for the connections between Virasoro complexity and Euler-Arnold type equations like the Korteweg-de Vries equation, which were pointed out in~\cite{Caputa:2018kdj,Erdmenger:2020sup}, but not followed up upon. We think that our results should add some clarity to the questions of \textit{which} Euler-Arnold type equations might arise as good complexity measures, to what kind of complexity respectively cost function they would correspond, and what geometric methods or objects might be relevant for their study. 

To be more specific, we reconsider the setting of \cite{Caputa:2018kdj,Erdmenger:2020sup}, which is a generic CFT$_{1+1}$ on a circle parametrized by the coordinate $\sigma$ and unitary transformations $U$ generated using the left- or right-moving component of the energy-momentum tensor $T(\sigma)$. As we review in section~\ref{sec:Conftrans}, such transformations can be thought of as diffeomorphisms of the circle defining a constant time slice in our theory. In~this setting the operators ${\cal O}_{I}$ defining circuits~\eqref{eq.circuit} are $T(\sigma)$, where $\sigma$ plays the role of the index $I$ and summation over $I$ is replaced by an appropriate integral. The key guiding principle behind our approach to this problem, as elucidated already in our earlier paper~\cite{Flory:2020eot}, is the requirement that the considered cost functions allow for a well-posed initial value problem between two arbitrary local conformal transformations. This leads us to consider in section~\ref{sec::FSmetric} the Fubini-Study distance in the setting in which the reference state $|\mR\rangle$ is given by an energy eigenstate~$|h\rangle$ corresponding to a primary or a quasiprimary operator of chiral dimension $h$ and the target state $|\mT\rangle$ is chosen to be $|\mT\rangle = U |h\rangle$. An important subtlety of this construction lies in the coincident point singularity of the two-point function of $T(\sigma)$, which in section~\ref{sec::diffreg} we regulate using differential regularization~\cite{Freedman:1992tz,FREEDMAN1992353,Latorre:1993xh,Erdmenger:1996yc}. The Fubini-Study metric leads to an intricate optimization problem in which geodesic equations are integro-differential equations and we discuss it in section~\ref{sec::affine}. Despite their complicated form, we were able to solve them in a perturbative manner around identity transformation, as outlined in section~\ref{sec::approx}. Subsequently, in section~\ref{sec::geodesics} we use these solution to elucidate properties of the underlying Fubini-Study geometry, in particular its sectional curvature. As we alluded earlier, our considerations of the Fubini-Study distance have led us to look at more general cost functions realizing Euler-Arnold equations of motion and this aspect of our work is discussed in section~\ref{sec::EA}. In this way we encapsulate both the state and circuit complexity associated with local conformal transformations in a single conceptual framework. We conclude with section~\ref{sec::conc} containing a summary and outlook. In appendices, we discuss more technical aspects of our work: null circuits representing global conformal transformations for the Fubini-Study distance when $h = 0$ (appendix~\ref{sec::SL2R}), the structure of our iterative solution for the Fubini-Study problem in the simplifying limit $h \gg c$ (appendix~\ref{sec::effingprimes}) and a way of dealing with metric degeneracies needed to define sectional curvature (appendix~\ref{sec::inverse}).


\section{Conformal transformations}
\label{sec:Conftrans}

In order to make our paper more self-contained and establish our notation, in this section we briefly review conformal transformations and the Virasoro group following the outline of the preceding works \cite{Caputa:2018kdj,Erdmenger:2020sup} and the excellent introduction~\cite{Oblak:2016eij}. 
For this, we consider the unit circle $S^1$ parametrized by an angular coordinate $\sigma\in [0,2\pi[,$ with $\sigma\sim \sigma+2\pi$. As we already mentioned in section~\ref{sec::Intro}, the circle is to be viewed as a spatial slice of a spacetime in which a CFT$_{1+1}$ lives. Orientation preserving diffeomorphisms on the circle are now smooth invertible maps of the form
\begin{align}
\label{eq.defstofs}
f: \sigma\rightarrow f(\sigma)
\end{align}
that map the circle to itself, hence they need to satisfy the periodicity condition $f(\sigma+2\pi)\sim f(\sigma)+2\pi$. Invertibility demands that $f'(\sigma)>0$ for any $\sigma$. These maps form the group of orientation preserving diffeomorphisms on the circle $S^1$, $Diff^{+}(S^1)$, with the group operation
\begin{align}
(f_1\cdot f_2)(\sigma)\equiv f_1\circ f_2(\sigma)=f_1(f_2(\sigma)).
\end{align}
The inverse of a group element is simply the inverse function, and the identity element is the map $f(\sigma)=\sigma$.

The Virasoro group is obtained as the central extension of the group of diffeomorphisms on the circle, thus an element in the Virasoro group is given by a function $f(\sigma)$ and a number $\alpha\in \mathbb{R}$. The rules for group multiplication are
\begin{align}
(f_1(\sigma),\beta)\cdot(f_2(\sigma),\alpha)=(f_1\circ f_2,\alpha+\beta+\mathfrak{C}(f_1,f_2)).
\end{align}
Herein, $\mathfrak{C}(f_1,f_2)$ is the Bott-cocycle which is explicitly given, for example, in \cite{Oblak:2016eij,Oblak:2017ect}.

The next step is to write circuits of the form \eqref{eq.circuit} in the language of the Virasoro group. As explained in \cite{Caputa:2018kdj}, for a path (or \textit{circuit}) of the form $f(\tcs)$ on the space of diffeomorphisms of the circle, parametrised by $\tau$, the~Hermitian generator $Q(\tau)$ obeys the relation
\begin{align}
&e^{-i\,Q(\tau)d\tau}U(\tau)\equiv U(\tau+d\tau)
\quad \Rightarrow \quad -i\,Q(\tau)=\dot{U}\,(\tau)\, U^{-1}(\tau),
\end{align}
and takes the form
\begin{equation}
\label{eq.Qcft}
Q(\tau) = \int_{0}^{2 \pi} \frac{d\sigma}{2 \pi} \, T(\sigma)\,\epsilon(\tau,\sigma).
\end{equation}
Here, we have defined
 \begin{align}
 \epsilon(\tau,f(\tcs))\equiv\dot{f}(\tcs).
 \label{epsilonf}
 \end{align}
We ignore here the real number component $\alpha(\tau)$ of the group element, as when acting on a state it will be associated with a complex phase, and it is generally assumed that complex phases should not play a role in studies of complexity.  We will however return to a more complete discussion in section \ref{sec::EA}. 
As we are assuming only one copy of the Virasoro algebra, $T(\sigma)$ can be interpreted as the left or right-moving component of the stress-energy tensor operator of a CFT$_2$ and $\epsilon(\tcs)$ is an element of the Lie-algebra.
The physical interpretation of \eqref{eq.Qcft} and \eqref{epsilonf} is that the infinitesimal operator $Q(\tau)\,d\tau$ generates an infinitesimal layer of the circuit \eqref{eq.circuit} at parameter $\tau$, which transforms the diffeomorphism $f(\tcs)$ to $f(\tau+d\tau,\sigma)$. In connection with~\eqref{eq.deflayergen}, one can see now explicitly what we have already stated in section~\ref{sec::Intro}, i.e. ${\cal O}_{I}$ can be taken to be $T(\sigma)$, the index $I$ is then the coordinate~$\sigma$, the sum $\sum_{I}$ becomes the integral $\int_{0}^{2 \pi} \frac{d\sigma}{2  \pi}$ and the velocity $Y^{I}(\tau)$ is nothing else than $\epsilon(\tau,\sigma)$.

As in~\cite{Caputa:2018kdj,Erdmenger:2020sup}, we assume for most of our paper that the circuit \eqref{eq.circuit}, with the generator now explicitly given in~\eqref{eq.Qcft}, acts on a reference state $|\mR\rangle$ such that we obtain a path on the space of states $|\psi(\tau)\rangle$ of the form~\eqref{psioftau}. Later, just as in~\cite{Caputa:2018kdj,Erdmenger:2020sup}, we will explicitly choose $|\mR\rangle$ to be an energy eigenstate in the CFT$_{1+1}$ on a cylinder. The energy eigenstates correspond then, via the operator-state correspondence, to primaries or quasiprimaries of the chiral dimension~$h$. We will denote such $|\mR\rangle$ by $|h\rangle$. The target state~$|\mT\rangle$, on the other hand, is going to be given by a unitary transformation realizing~\eqref{eq.defstofs} acting on $|h\rangle$ and will be a superposition of $|h\rangle$ and states lying in the respective Virasoro tower.

Starting from the next section, we will consider the Fubini-Study complexity encapsulated by~\eqref{eq.costFS}, which will require us to evaluate the two-point function $\left\langle\psi(\tau)|Q^2(\tau)|\psi(\tau)\right\rangle$ of the generator $Q(\tau)$ in the state~$\left|\psi(\tau)\right\rangle$ both \textit{at parameter $\tau$}, as well as the associated one-point function $\left\langle\psi(\tau)|Q(\tau)|\psi(\tau)\right\rangle$. We can reduce them to correlation functions in the state $\hrr$ by introducing the generator \cite{Caputa:2018kdj}
\begin{align}
\tilde{Q}(\tau) = U(\tau)^{\dagger} \, Q(\tau)\, U(\tau),
\label{tildeQ}
\end{align}
such that evidently $\hll\tilde{Q}^2(\tau)\hrr=\left\langle\psi(\tau)|Q^2(\tau)|\psi(\tau)\right\rangle$, and similarly for the one-point function.

The applications of the operator $U(\tau)$ in \eqref{tildeQ}, using  \eqref{eq.Qcft}, essentially cause a local conformal transformation of the stress-energy tensor under the integral. Using the well-known transformation law, we can write \cite{Caputa:2018kdj}
\begin{equation}
\tilde{Q}(\tau) = \int_{0}^{2 \pi} \frac{d\sigma}{2 \pi} \, \frac{\dot{f}(\tcs)}{f'(\tcs)}\left(T(\sigma)-\frac{c}{12}\{f(\sigma),\sigma\}\right)
\label{tildeQ2}
\end{equation}
where $\{f(\sigma),\sigma\}$ is the Schwarzian derivative. The Schwarzian will only contribute a complex phase to the circuit~\eqref{eq.circuit}. This contribution to $U(\tau)$ should not be important for any good definition of state complexity, as states themselves are defined up to an overall phase. Since unitary operators are meant to act on states, an analogous argument regarding phases applies also to any good definition of unitary complexity. Later, when we calculate \textit{connected} two-point functions of $\tilde{Q}$, which enter directly the Fubini-Study cost function~\eqref{eq.costFS}, the Schwarzian term will automatically cancel. 


\section{Complexity via the Fubini-Study metric}
\label{sec::FSmetric}

\subsection{Derivation}

While we have already introduced the Fubini-Study cost function as~\eqref{eq.costFS}, let us start by reminding readers the precise link between this formula and the Hilbert space distance~\cite{geomquantstates}. For a family of states like our $|\psi(\tau)\rangle =U(\tau)\, |\mR\rangle$, see~\eqref{psioftau}, the overlap between the nearby states is given by
\begin{align}
|\langle \psi(\tau)|\psi(\tau+d\tau)\rangle|\approx 1-G_{\tau\tau}(\tau)d\tau^2+\mO(d\tau^3)
\label{fsdefinition}
\end{align}
where $G_{\tau\tau}$ is the fidelity susceptibility \cite{PhysRevE.76.022101,2015JPCM...27t5601Y}, see also \cite{doi:10.1142/S0217979210056335,geomquantstates,Braunstein:1994zz} for background and \cite{Nozaki:2012zj,MIyaji:2015mia} for its uses in holography. 
The fidelity susceptibility, which in this case is equivalent to the Fubini-Study-metric, reads~\cite{Caputa:2018kdj}
 \begin{align}
 G_{\tau\tau}(\tau)=\frac{1}{2}\left(\langle Q^2 \rangle-\langle Q \rangle ^2\right),
 \label{FSmetric}
 \end{align}
where the expectation values are evaluated in the state $|\psi(\tau)\rangle$. This means that the Fubini-Study-metric is defined via the (connected part of the) two-point function $\langle Q^2 \rangle$. 
 Since one has a freedom to pick the way one parametrizes circuits at will, the invariant combination along the whole circuit from the reference state $|\psi(0)\rangle =|\mR\rangle$ to the target state $|\psi(1)\rangle =|\mT\rangle$, up to an unimportant normalization, is proportional to $\int_{0}^{1} d\tau \sqrt{G_{\tau\tau}}$. Direct evaluation makes this integral equal to the Fubini-Study cost function~\eqref{eq.costFS}. Since there resides a connected correlator under the square root, the complex phase contributing to the one point functions of $Q$ cancels, as anticipated.

To proceed, we will need the connected two-point function of $\tilde{Q}$ in the state $|\mR\rangle=|h\rangle$, which takes the form
\begin{align}
&\hl \tilde{Q}\tilde{Q}\hr -\hl \tilde{Q}\hr \hl\tilde{Q}\hr 
\label{2ptQ}
\\
&=\int_{0}^{2\pi} \frac{d\sigma d\kappa}{4\pi^2}  \frac{\dot{f}(\tcs)}{f'(\tcs)}\frac{\dot{f}(\tck)}{f'(\tck)}  
\left(\hl T(\sigma)T(\kappa) \hr - \hl T(\sigma) \hr \hl T(\kappa) \hr \right)
\nonumber
\end{align}
and indeed one sees that the phase related to the Schwarzian dropped out. The energy-momentum tensor $T(\sigma)$ can be decomposed in terms of Virasoro-generators $L_n$ via
\begin{subequations}
\label{VirasoroAlgebra}
\begin{align}
T(\sigma)&=\sum_{n\in\mathbb{Z}} \left(L_n-\frac{c}{24}\delta_{n,0}\right)e^{-in\sigma},
\label{T}
\\
[L_m,L_n]&=(m-n)L_{m+n}+\frac{c}{12}(m^3-m)\delta_{m+n,0}.
\end{align}
\end{subequations}
The energy eigenstate $\hrr$ acting here as the reference state satisfies the following relations in terms of Virasoro generators
\begin{subequations}
\begin{align}
L_n \hrr &=0 \text{  for  } n>0,  \hll L_n=0 \text{  for  } n<0,
\label{h1}
\\
L_0\hrr&= h\hrr,
\label{h2}
\\
\hl\mathbbm{1}\hr&=1.
\label{h3}
\end{align}
\end{subequations}
For this state, besides $\hl T(\sigma)\hr=h-\frac{c}{24}$, we also have the explicit result for the stress-energy two-point function provided, for example, in~\cite{Datta:2019jeo},
\begin{align}
\hl T(\sigma)T(0)\hr&=\left(h-\frac{c}{24}\right)^2+\frac{c}{32\sin(\sigma/2)^4}-\frac{h}{2\sin(\sigma/2)^2}.
\label{2pt}
\end{align}
Combining all the expressions, the Fubini-Study cost associated with the circuit~\eqref{eq.costFS} takes the form\footnote{We could have used \eqref{T} to phrase \eqref{2ptQ} and hence \eqref{complexity} in terms of a Fourier series instead of an integral over the coordinates~$\sigma,\kappa$. However, even if the Fourier series of $f(\tcs)$ is known, there is no general formula for the Fourier series of $\dot{f}(\tcs)/f'(\tcs)$, and consequently we prefer to work in this paper with expressions in terms of the smooth functions $f(\tcs)$ instead of Fourier modes. However, we will effectively make use of a (terminating) Fourier series expression in section~\ref{sec::geodesics}.  }
\begin{align}
\mathrm{L}_{\mathrm{FS}}=\int_0^1 \frac{d\tau}{2\pi} \sqrt{\iint_0^{2\pi}d\sigma 
d\kappa  \frac{\dot{f}(\tcs)}{f'(\tcs)}\frac{\dot{f}(\tck)}{f'(\tck)}  
\left(\frac{c}{32\sin((\sigma-\kappa)/2)^4}-\frac{h}{2\sin((\sigma-\kappa)/2)^2}\right)
 }.
\label{complexity}
\end{align}

\subsection{The geodesic analogy}
\label{sec::ga}

The problem of finding the circuits extremising \eqref{complexity} corresponds to the geodesic problem in uncountably-infinite dimensions, where we use the continuous  coordinates $\sigma,\kappa\in \left[0,2\pi\right[$ instead of discrete indices and make what we call the \textit{geodesic analogy} replacements
 \begin{align}
 \sum_\sigma &\leftrightarrow \int d\sigma,
 \label{geoan1}
 \\
X^\sigma(\tau) &\leftrightarrow f(\tau,\sigma),
 \label{geoan2}
\\
g_{\sigma\kappa} &\leftrightarrow \frac{\Pi(\sigma-\kappa)}{f'(\tcs)f'(\tck)}.
 \label{geoan3}
 \end{align}
Here, we have written 
\begin{align}
\Pi(\sigma-\kappa)=\hl T(\sigma)T(\kappa) \hr - \hl T(\sigma) \hr \hl T(\kappa) \hr   
\end{align}
as shorthand-notation for the energy-momentum-tensor two-point function which appears as integration kernel in \eqref{complexity}. We will mostly keep this abstract notation and derive results for a general choice of $\Pi$ whenever possible. We will however always assume that $\Pi$ is only a function of $\sigma-\kappa$, and that it is symmetric under swapping the coordinates $\sigma\leftrightarrow\kappa$.

\subsection{Differential regularisation}
\label{sec::diffreg}

Clearly, the two-point function \eqref{2pt} has poles at coincident insertion points and we will need to implement a regularisation-scheme in order to give the integral in \eqref{complexity} a well-defined finite value\footnote{
There is a superficial similarity between our expression for fidelity susceptibility~\eqref{fsdefinition} and the calculations presented in a different context in \cite{MIyaji:2015mia,Bak:2015jxd,Trivella:2016brw,Bak:2017rpp}, insofar as the integration over a two-point function plays a role. Note however that in \cite{MIyaji:2015mia,Bak:2015jxd,Trivella:2016brw,Bak:2017rpp} a two-point function $\langle \mO(t,x)\mO(t',x')\rangle$ is integrated over the space and time coordinates of both operator insertions, in such a manner that $|t-t'|\geq2\delta$ for some ultraviolet (UV)-cutoff $\delta$, such that the pole of the two-point function is not in the integral domain. The result obtained there will then include a leading order divergent term in $\delta$.  
This is different from~\eqref{2ptQ} where the chiral component of the stress-energy tensor $T$ is dependent on one coordinate $\sigma$ or $\kappa$, and the pole of the two-point function lies in the integration domain. Nevertheless, based on the results of \cite{Flory:2018akz,Belin:2018bpg}, we expect to find UV-finite expressions for complexity as the reference state in the UV has the same behaviour as the target state. Furthermore, somewhat similar formulas to \eqref{2ptQ} appear in the context of chaos in CFTs$_{1+1}$ in \cite{Haehl:2018izb}.}.
There may be different ways to accomplish this, but in this paper we will settle on the method of \textit{differential regularisation}, see \cite{Freedman:1992tz,FREEDMAN1992353,Latorre:1993xh,Erdmenger:1996yc}. 
In this method, one essentially phrases the divergent terms of the two-point function~\eqref{2pt} as derivatives of terms with milder, integrable singularities and then carries the derivatives over onto the test-functions via integration by parts. In such an approach, one treats the two-point function as a distribution, against which test-functions are integrated, such that derivatives of the distribution are \textit{defined} via integration by parts.

To this end, the $h$-dependent part of $\Pi$ is written as
\begin{align}
\label{eq.hpartdifreg}
-\frac{h}{2\sin((\sigma-\kappa)/2)^2}=-h\,\partial_\sigma\partial_\kappa\log\left[\sin\left((\sigma-\kappa)/2\right)^2\right]
\end{align}
where $\log\left[\sin\left((\sigma-\kappa)/2\right)^2\right]$ has a pole at $\sigma=\kappa$ which is mild enough to be integrated over when the integration-kernel is multiplied with a smooth test-function, yielding a finite and real-valued result.

The next term to be taken care of is the $c$-dependent part of the two-point function. For this, we rewrite it as
\begin{align}
&\frac{c}{32\sin((\sigma- \kappa)/2)^4}-\frac{c}{48\sin((\sigma- \kappa)/2)^2}+\frac{c}{48\sin((\sigma- \kappa)/2)^2} \nonumber
\\
&=\frac{c}{96}\left(2+\cos(\sigma- \kappa)\right)\csc\left((\sigma- \kappa)/2\right)^4+\frac{c}{48\sin((\sigma- \kappa)/2)^2}
\end{align}
We already know how to (differentially) regularise the second term via~\eqref{eq.hpartdifreg}. For the first term, we find
\begin{align}
\frac{c}{96}\left(2+\cos(\sigma- \kappa)\right)\csc\left((\sigma- \kappa)/2\right)^4= - \frac{c}{24}\partial_\sigma^2\partial_\kappa^2\log\left[\sin((\sigma- \kappa)/2)^2\right].
\end{align}
Then, using these expressions and applying integration by parts with impunity, we arrive at
\begin{subequations}
\begin{align}
&\iint_0^{2\pi}d\sigma 
		d\kappa  \frac{\dot{f}(\tcs)}{f'(\tcs)}\frac{\dot{f}(\tck)}{f'(\tck)}  
\left(\frac{c}{32\,\sin((\sigma-\kappa)/2)^4}-\frac{h}{2\,\sin((\sigma-\kappa)/2)^2}\right)
				\label{diffreg-1}
			\\
		&\overset{!}{=}\iint_0^{2\pi}d\sigma d\kappa	\log\left(\sin\left(\frac{\sigma-\kappa}{2}\right)^2\right)
						\label{diffreg}
		\\
		&\times\left(-\frac{c}{24}\partial_\sigma^2 \frac{\dot{f}(\tcs)}{f'(\tcs)}\partial_\kappa^2\frac{\dot{f}(\tck)}{f'(\tck)}  +\left(\frac{c}{24}-h\right)\partial_\sigma \frac{\dot{f}(\tcs)}{f'(\tcs)}\partial_\kappa\frac{\dot{f}(\tck)}{f'(\tck)}  \right).
\nonumber
\end{align}
\end{subequations}

\subsection{Degeneracy of the metric}
\label{sec::deg}

The differential regularisation has the important consequence that our metric will not be positive definite, instead there will be certain directions in the tangent space which will be null. This is easy to see by setting $f(0,\sigma)=\sigma, \dot{f}(0,\sigma)=1, \Rightarrow \partial_\sigma \frac{\dot{f}(0,\sigma)}{f'(0,\sigma)}=0$ and \eqref{diffreg} obviously vanishes. Let us study the null directions of the tangent space at the point $f(\sigma)=\sigma$ ($\Rightarrow f'(\sigma)=1$) in the space of maps more closely. For a tangent vector $\dot{f}(0,\sigma)\equiv v(\sigma)=\sin(n\sigma+\phi)$, $n\in\mathbb{Z}$, $0\leq\phi<2\pi$, we find the norm 
\begin{align}
||v||^2\equiv \iint_0^{2\pi}d\sigma 
		d\kappa \ v(\sigma)\,v(\kappa) \, \Pi(\sigma-\kappa)
		=\frac{1}{12} \pi ^2 n \left(24\, h+c \left(n^2-1\right)\right).
\label{normv}
\end{align}
Clearly, for generic $c,h>0$ this equals zero for $n=0$ (i.e.~the null direction which we have identified above), and is otherwise positive, indicating a positive-semidefinite degenerate metric. The null-direction $n=0\Rightarrow v(\sigma)=const.$ corresponds to the $U(1)$ subgroup of the Virasoro-group which is generated by $L_0$ alone and acts on the state $\hrr$ by giving it at most a complex phase, as $\hrr$ is an eigenstate of this generator. In light of our earlier discussion about phases, it is natural that our complexity definition assigns zero distance to such an operation.

Let us next look at the special case $h=0$, where the reference state is the vacuum state $\left|0\right\rangle$ which is invariant under the $PSL(2,\mathbb{R})$-subgroup of the Virasoro-group generated by $L_{-1},L_0$ and $L_1$, i.e. half of the group of global conformal transformations. Indeed, for $h=0$, \eqref{normv} can be zero for $n=0$ and $n=\pm1$. Hence, just as in \cite{Erdmenger:2020sup}, this case is special for our notion of complexity on the Virasoro-group, and we will generically assume $h\neq0$ in the following and delegate more details about the $h = 0$ case to appendix~\ref{sec::SL2R}.  
Equation \eqref{normv} could also be zero for $n=0$ and $n^2=1-24\frac{h}{c}$, which would correspond to the Virasoro-subgroups $SL^{(n)}(2,\mathbb{R})$ generated by $L_0,L_{-n},L_n$~\cite{Oblak:2016eij,Erdmenger:2020sup}. Since this case requires $h<0$ inconsistent with unitary CFT$_{1+1}$, it is unphysical and we dismiss it, as it was also done in a related context in~\cite{Erdmenger:2020sup}.

Which operators only generate phases depends of course on the choice of reference state, as Fubini-Study distance is a notion of complexity defined on the space of states directly. 
We hence see that all the stabilizer subgroups of the Virasoro group that already played a prominent role in \cite{Erdmenger:2020sup} appear in our approach as degeneracies of the metric for specific choices of $h$. Thus, circuits generated by these groups will automatically be \textit{null} in our framework, i.e.~will be assigned zero complexity cost, while they however did receive non-vanishing cost under the framework of \cite{Erdmenger:2020sup} and earlier work~\cite{Caputa:2018kdj}.


 \section{Equations of motion and their solutions}
 \label{sec::affine}
 
\subsection{Variational Principle}
\label{sec::VIP}
 
When deriving the geodesic equation in ordinary Riemannian geometry, it is a well known trick based on reparametrization invariance that we can use either the Lagrangian $\mL_1=\sqrt{g_{\mu\nu}\dot{X}^\mu\dot{X}^\nu}$ or $\mL_{sq}=\mL_1^2=g_{\mu\nu}\dot{X}^\mu\dot{X}^\nu$. Specifically, it can be shown that the Euler-Lagrange-equations following from a variation of $\mL_{sq}$ are equivalent to those following from $\mL_1$ if $\frac{d\mL_{sq}}{d\tau}\propto\frac{d\mL_1}{d\tau}=0$, i.e.~for affine parametrization. More specifically, every solution to $\mL_{sq}$ is affinely parametrized and is a solution to $\mL_1$, and every affinely parametrized solution to $\mL_1$ (but not non-affine ones) is a solution to $\mL_{sq}$. 
The benefit in using $\mL_{sq}$ instead of $\mL_1$ is that due to the absence of the square-root, the equations of motion will take on a simpler form. 
The reason for the important role of affineness in this is that both Lagrangians do not depend explicitly on the variable $\tau$, and hence are independent under shifts of the form $\tau\rightarrow \tau+\delta \tau$. Due to Noether's theorem, there is hence a conserved quantity (energy), that takes the form $\mQ=-\mL+\frac{\partial\mL}{\partial \dot{X}^\mu}\dot{X}^\mu$. For $\mL_1$, $\mQ=0$ due to reparametrization invariance, while for $\mL_{sq}$, $\mQ\sim\mL_{sq}$, implying affineness. Similarly, the functional \eqref{complexity} is invariant under reparametrizations due to the presence of the square-root, and we will likewise work with the squared Lagrangian $\mathrm{L}_{sq}$ in this section for convenience. In order to do this, we introduce the functional\footnote{To be clear about terminology, we recognize of course that $\mathrm{L}_{sq}$ is the Lagrangian while $\mL$ is a Lagrange density in \eqref{complexityf2}. }
  \begin{align}
	\mathrm{L}_{\mathrm{FS},2}&=	\int_0^1 \frac{d\tau}{2\pi} \mathrm{L}_{sq}[f,f',\dot{f}] = 	\int_0^1 \frac{d\tau}{2\pi} \iint_0^{2\pi}d\sigma 
	d\kappa\ \mL[f,f',\dot{f}]
	\\  	
	&
	=	\int_0^1 \frac{d\tau}{2\pi} \iint_0^{2\pi}d\sigma 
	d\kappa  \Pi(\sigma-\kappa)  
	\frac{\dot{f}(\tcs)}{f'(\tcs)} \frac{\dot{f}(\tck)}{f'(\tck)}
	\label{complexityf2}
\end{align}
instead of \eqref{complexity}, which is now formally written as 
  \begin{align}
	\mathrm{L}_{\mathrm{FS}}&=	\int_0^1 \frac{d\tau}{2\pi} \sqrt{\mathrm{L}_{sq}[f,f',\dot{f}]} = 	\int_0^1 \frac{d\tau}{2\pi}\sqrt{ \iint_0^{2\pi}d\sigma 
	d\kappa\ \mL[f,f',\dot{f}]}.
	\label{complexityf1}
\end{align}
Extremising the squared Lagrangian yields the following integro-differential equation (IDE) of motion
 \begin{align}
 \int_0^{2\pi}d\sigma\Bigg(-\Pi(\sigma-\kappa)\frac{d}{d\tau}\left(\frac{\dot{f}(\tcs)}{f'(\tcs)f'(\tck)}\right)+\frac{\dot{f}(\tcs)}{f'(\tcs)}\partial_\kappa\left(\Pi(\sigma-\kappa)\frac{\dot{f}(\tck)}{f'(\tck)^2}\right)\Bigg)
 \equiv 0,
 \label{EOMf2}
 \end{align}
which is the main equation studied in this paper and our earlier work~\cite{Flory:2020eot}.  
Three additional comments about~\eqref{EOMf2} are in order. Firstly, we see that~\eqref{EOMf2} is formally a second order equation in $\tau$-derivatives. Consequently, solutions will depend on chosen boundary conditions, which might come in form of an initial profile of the diffeomorphism and its first derivative with respect to the circuit parameter ($f(0,\sigma),\dot{f}(0,\sigma)$) or an initial and a final profile of the diffeomorphism ($f(0,\sigma),f(1,\sigma)$). The latter is what we generally imagine to be relevant for a complexity setting, where we are interested in the optimal circuit that implements a given operation starting from the identity operator. 
Secondly, however, the second order derivative with respect to the circuit parameter appears under the integral, so even with given initial conditions $f(0,\sigma),\dot{f}(0,\sigma),$ we could not \textit{directly} calculate the next time-step in an Euler-like numerical scheme, because we cannot solve immediately for $\ddot{f}(0,\sigma)$ as a function of the initial conditions. 
Thirdly, the differential regularisation leads to terms of the form\footnote{In equation \eqref{diffreg}, we had carried out differential regularisation symmetrically by integration by parts in both $\sigma$ and $\kappa$. In the EOM \eqref{EOMf2}, there is only the integral over $\sigma$ left, hence differential regularisation is carried out only via derivatives in $\sigma$, which thus reach up to fourth order.}  $\ddot{f}''''(\tcs)$ and $\ddot{f}''(\tcs)$ besides $\ddot{f}(\tcs)$, which seems unusual based on our intuition for PDEs and again would prevent us from straightforwardly implementing an Euler-like numerical scheme. We will come back to these features in section~\ref{sec::U1gauge} and appendix~\ref{sec::inverse}.

Finally, let us elaborate on several features related to~\eqref{EOMf2} being an IDE. The reason that we obtained an~IDE as a mathematical description of our problem is easy to understand from the geodesic analogy introduced in section \ref{sec::ga}. Essentially, we are dealing here with a geodesic equation 
\begin{align}
g_{\mu\nu}\ddot{X}^\mu+\frac{1}{2}\left(g_{\alpha\nu,\beta}+g_{\beta\nu,\alpha}-g_{\alpha\beta,\nu}\right)\dot{X}^\alpha\dot{X}^\beta=0.
\label{degenerategeodesics}
\end{align}
in which a summation over indices has been replaced by an integration over a continuous variable as in \eqref{geoan1}. Note that we have deliberately not brought \eqref{degenerategeodesics} to its "usual" form by contracting with the inverse metric $g^{\rho\nu}$ because such an inverse does not exist in our case due to null directions discussed in~\ref{sec::deg}. We revisit this feature in appendix \ref{sec::inverse}. Furthermore, while in general IDEs do find uses in mathematical physics~\cite{lakshmikantham1995theory} and, what might be interesting in light of the currently ongoing SARS-CoV-2 pandemic, also in modeling the spread of diseases \cite{10.2307/2653135}, they may not feel very natural to most general relativists and quantum field theorists. In these fields one more often deals with partial differential equations and the reason for it is that they respect a notion of locality, such as interactions between only nearest neighbour particles or propagation of information at finite speeds. Such locality requirements need not to be there in definitions of complexity, where we may want to assign a cost to a layer of a quantum circuit~\eqref{eq.circuit} in such a way that the presence of a gate at one point of the layer may affect how we account for a gate at any other position of the layer. In this sense, one should not be surprised by the appearance of an IDE in the complexity context.

\subsection{Trivial solutions, circular fibres}
\label{sec::fibres}

 Let us begin by looking at simple solutions of \eqref{EOMf2}. The trivial solution is of course $f(\tcs)=f(\sigma)=\mathrm{const}$, i.e.~an~analogue of a trivial geodesic "solution" $\dot{X}^\mu=0$. More interestingly, for general $\Pi$ we find it easy to verify that $f(\tcs)=\sigma+\tau$ is a solution. These are geodesics which start at the identity $f(0,\sigma)=\sigma$ and then just get a $\tau$-dependent translation. Due to the periodicity of the circle, $\sigma\sim\sigma+2\pi$, these geodesics are periodic in $\tau$ with period $2\pi$, after which they return to the starting point. These solutions start from identity and there acquire a natural generalization if one starts from other group element, namely
\begin{align}
f(\tcs)&=F(\sigma+\tau)
\end{align}
for any suitable function $F(\sigma)$. 
Due to the degeneracy of the metric described in section \ref{sec::deg} these solutions are null-geodesics for our concrete choice of $\Pi$.

This can be understood as a fibration of the space of diffeomorphisms on the circle, with the trivial circular geodesics being the fibres. The base space can, for example, be defined by all maps $f(\sigma)$ which satisfy the condition $f(0)=0$, as discussed later in section \ref{sec::U1gauge} and especially figure \ref{fig::gauge}. See \cite{Oblak:2016eij} for more details about the topology of the group $Diff^+(S^1)$.

\subsection{Symmetries and conserved charges}

The system \eqref{complexity} and its equivalent form \eqref{complexityf2} exhibit a number of interesting continuous symmetries, for which we can derive conserved charges via the usual Noether procedure. We will go through these symmetries one by one in this section. 

\subsubsection{Affine shifts}
 
While the functional \eqref{complexityf2} has lost the full $\tau$-reparametrization invariance of \eqref{complexity}, it is still invariant under affine shifts of the form $\tau\rightarrow \tau+\delta \tau$. As already discussed in section~\ref{sec::VIP}, the conserved quantity is
\begin{align}
\mQ=\mathrm{L}_{sq}.
\label{conEnergy}
\end{align}
This, again, will ensure affineness on all solutions. In fact, we can left multiply the equations of motion \eqref{EOMf2} by $\int d\kappa( \dot{f}(\tck)\times...)$ to show explicitly that this quantity will be conserved along any solution.

\subsubsection{Conformal symmetry}
\label{sec::conformal}

A different symmetry of the system would be
\begin{align}
 f\rightarrow F(f)\equiv f+\delta g(f),
 \label{conformalsymmetry}
\end{align}
with arbitrary maps $F(f)$ respectively $g(f)$, subject to suitable periodicity conditions:
\begin{subequations}
\begin{align}
F(\sigma+2\pi)&\equiv F(\sigma)+2\pi,
\\
F(\sigma)&\equiv\sigma+\delta g(\sigma),
\\
\Rightarrow\delta g(\sigma+2\pi)&=\delta g(\sigma).
\end{align}
\end{subequations}
This is clearly related to the group of diffeomorphisms on the circle discussed in section \ref{sec:Conftrans}, and is trivially a symmetry of the Lagrangian as it maps $\frac{\dot{f}(\tcs)}{f'(\tcs)}$ to itself
\begin{align}
\frac{\dot{f}(\tcs)}{f'(\tcs)}\rightarrow \frac{F'(f(\tcs))}{F'(f(\tcs))} \frac{\dot{f}(\tcs)}{f'(\tcs)}.
\end{align} 
Essentially, this just shows that the inner product \eqref{complexity} is left-invariant. Note that this kind of symmetry also played a pivotal role in \cite{Erdmenger:2020sup}.

It is important to point out that it is \textit{because of this symmetry} that the functional \eqref{complexityf2} yields a well-defined distance measure on the group of diffeomorphisms on the circle (respecting the group properties), \textit{not} because of the specific choice of $\Pi$ as a two-point function of a CFT. 
This means that we may also study other choices of $\Pi$, such as $\Pi(\sigma-\kappa)=\delta(\sigma-\kappa)$  or $\Pi(\sigma-\kappa)=\delta''(\sigma-\kappa)$. We will come back to this in section~\ref{sec::EA}.

Equation \eqref{conformalsymmetry} leads to the conserved quantity
 \begin{align}
 \mQ_{\delta g}= 2\iint_0^{2\pi}d\sigma \,d\kappa\, \Pi(\sigma-\kappa)\left(\frac{\dot{f}(\tcs)}{f'(\tcs)\,f'(\tck)}\right)\delta g(f(\tck)).
\label{Qg}
 \end{align}
Again, this can also be derived directly from the equations of motion \eqref{EOMf2}, multiplying the whole expression from the left with $\int d\kappa( \delta g(f(\tck))\times...)$, and treating the $\partial_\kappa$ derivative with integration by parts. It should be noted that $\mQ_{\delta g}$ is an \textit{infinite} family of conserved charges, as it is conserved for \textit{any choice} of $\delta g$. Because of this arbitrariness of $\delta g$, the conservation of $\mQ_{\delta g}$ actually implies the EOMs \eqref{EOMf2}:
 \begin{align}
 EOMs=0\Leftrightarrow \frac{d}{d\tau}\mQ_{\delta g}=0\quad \forall \delta g.
 \end{align}
Does this mean that our system is \textit{Liouville-integrable} (see, for example,~\cite{Torrielli:2016ufi})? For that to hold, the set of conserved charges $\{\mQ_{\delta g}\}$ would have to be independent (which is true) and in \textit{involution}, i.e.~the Poisson bracket of any two of these charges has to vanish. This turns out \textit{not} the case for the following reason. As expected, it is easy to see that \eqref{conformalsymmetry} maps solutions of the EOMs to other solutions, as equation \eqref{EOMf2} essentially just acquires a prefactor $\frac{1}{F'(f(\tck))}$, as well as two terms which exactly cancel each other. How does such a map affect the conserved quantities? It is easy to see that under $f\rightarrow F(f)$,
 \begin{equation}
 \mathrm{L}_{sq}\rightarrow \mathrm{L}_{sq} \quad \mathrm{and} \quad \mQ_{\delta g}\rightarrow  \mQ_{\delta \hat{g}} 
 \end{equation}
 with
\begin{equation}\delta \hat{g}(f(\tck))\equiv \frac{\delta g(F(f(\tck)))}{F'(f(\tck))}\approx \delta g(f)+\delta g'(f)\delta k(f)-\delta g(f)\delta k'(f) \quad \mathrm{for} \quad F(f)=f+\delta k(f).
\end{equation}
So $\delta\hat{g}=\delta g$ if and only if $\delta k=\delta g$ in general. Thus, under a symmetry transformation corresponding to the charge $\mQ_{\delta k}$, the charge $\mQ_{\delta g}$ corresponding to a different symmetry will in general transform nontrivially, so their Poisson bracket cannot vanish. Due to absence of Liouville integrability, instead of trying to find action-angle coordinates to solve~\eqref{EOMf2}, in section~\ref{sec::approx} we will solve perturbatively for transformations that are close to each other.

\subsubsection{From a global $U(1)$ symmetry to a local $U(1)$ gauge freedom}
\label{sec::U1gauge}

An additional symmetry turns out to be shifts along the circle, $\delta\sigma=\mathrm{const}$, which cause
\begin{align}
f(\tau,\sigma)\rightarrow f(\tau,\sigma+\delta\sigma)\Rightarrow \mQ'= 2\iint_0^{2\pi}d\sigma\, d\kappa \, \Pi(\sigma-\kappa)\,\frac{\dot{f}(\tcs)}{f'(\tcs)}.
\end{align} 
This can be shown by using that $\Pi$ is a function of $\sigma-\kappa$, and hence 
\begin{align}
\partial_{\kappa}\Pi(\sigma-\kappa)=-\partial_{\sigma}\Pi(\sigma-\kappa),
\end{align} 
We can show explicitly that $\frac{d}{d\tau}\mQ'=0$ by left-multiplying the EOMs \eqref{EOMf2} with $\int d\kappa( f'(\kappa)\times...)$. This symmetry shifts a function around the circle,
hence due to the periodicity of $\sigma$, this is a (global) $U(1)$-symmetry.  

\begin{figure}[htbp]
	\centering
	\def\svgwidth{0.8\columnwidth}
\executeiffilenewer{gauge.svg}{gauge.pdf}%
{inkscape -z -D --file=gauge.svg %
--export-pdf=gauge.pdf --export-latex}%
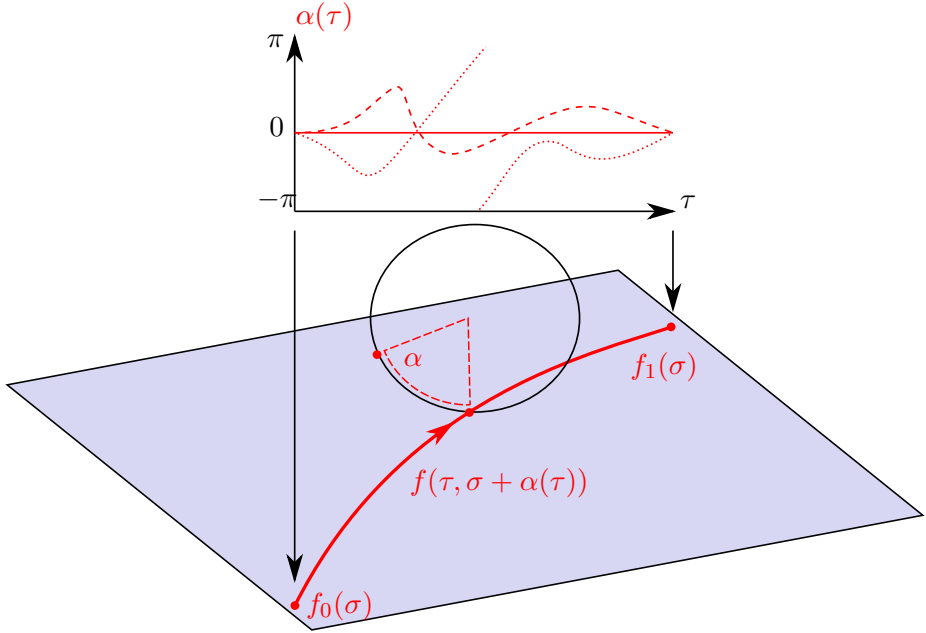%

	\caption{The space of allowed maps can be described via circular fibres over a base-space. This base space can be described as the space of allowed maps for which $f(0)=0$, while the points along the fibre have $f(\alpha)=0$. The circuit $f(\tau,\sigma)=f(\tau+\sigma)$	is trivially a geodesic for any $\Pi$ and winds around such a fibre as an uncontractible cycle, see section \ref{sec::fibres}. 
		Assume for a given $\Pi$, a geodesic circuit $f(\tau,\sigma)$ (solid red curve) connects two points $f_0(\sigma),f_1(\sigma)$ in the base space (shaded area). If the cycles are null, then this circuit can be wiggled along the circle by an arbitrary function $\alpha(\tau)$ (solid, dashed and dotted red lines above), and $f(\tau,\sigma+\alpha(\tau))$ will still be a geodesic of identical length -- the geodesic problem on the degenerate metric has a $U(1)$-gauge redundancy.  }
	\label{fig::gauge}
\end{figure}

In a case where we use differential regularisation, something interesting happens. When 
\begin{align}
    \Pi(\sigma-\kappa)=\partial_\sigma\partial_\kappa\tilde{\Pi}(\sigma-\kappa)
    \label{hatPI}
\end{align}
for some $\tilde{\Pi}$, integration by parts in $\kappa$ above clearly shows that $\mQ'=0$ also when $\delta \sigma$ depends on~$\tau$. This is because the circular fibres are now \textit{null} (see section \ref{sec::deg}), and the $U(1)$ shift symmetry becomes a \textit{gauge freedom}. If $f(\tau,\sigma)$ is a solution to the equations of motion, then so will be $f(\tau,\sigma+\alpha(\tau))$ for \textit{any} function $\alpha(\tau)$, as wiggling a geodesic along the circular null-direction does not incur any cost in terms of distance traversed\footnote{Circuits of the form $f(\tcs)=g(\sigma+\alpha(\tau))$ with arbitrary $\alpha(\tau)$ also appeared in \cite{Erdmenger:2020sup} as solutions to their equations of motion. However, there the associated cost was nontrivial.}. From the point of view of defining a distance measure, this wiggling of the geodesic along the circular fibres is hence a redundancy in our physical description of the system\footnote{
Let us note here that geodesic motion on degenerate metrics with a circular null-direction has also been studied in the context of an alternative to Kaluza-Klein unification of gravity and electromagnetism in \cite{Searight:2003vk,Searight:2017nav}, see also \cite{Stoica:2016bro}.}. See figure \ref{fig::gauge} for an illustration.

The presence of the gauge redundancy explains why when using differential regularisation the highest $\tau$-derivative in the EOMs can appear in the form of a combination of terms involving $\ddot{f}(\tcs),\ddot{f}''(\tcs)$ and higher spacial derivatives, as we already noted in section~\ref{sec::VIP}. In order to calculate a numerical solution to the EOMs via an Euler-like scheme, we would have to compute the second order time-derivative $\ddot{f}(0,\sigma)$ for given initial conditions $(f(0,\sigma),\dot{f}(0,\sigma))$. If we can only obtain a combination of $\ddot{f}(0,\sigma),\ddot{f}''(0,\sigma),...$ etc~as a function of the initial conditions, this gives us an ordinary differential equation for $\ddot{f}(0,\sigma)$ in $\sigma$ that has to be solved before the next numerical step in $\tau$ can be calculated. Solving this equation will require boundary conditions. Most of these boundary conditions can be fixed by requiring periodicity along the circle (e.g.~$\ddot{f}''(0,\sigma)=\ddot{f}''(0,\sigma+2\pi)$), but the last step yielding  $\ddot{f}(0,\sigma)$ will generally be ambiguous. This ambiguity corresponds to the freedom of choosing the function $\alpha(\tau)$, and
can be (gauge-) fixed by demanding a condition like $f(\tau,0)=0\ \forall \tau$, i.e.~demanding that the circuits lie in the base space. This would correspond to only considering the group $Diff_0^+(S^1)$ in \cite{Oblak:2016eij}.

As pointed out in section \ref{sec::deg}, in the special case $h=0$ the degeneracy of the metric is extended by the appearance of additional null-directions. Based on our understanding developed so far, we expect the $U(1)$-gauge redundancy to be enhanced to an $PSL(2,\mathbb{R})$ gauge-group for $h=0$. We confirm this explicitly in appendix \ref{sec::SL2R}.

\subsection{Iterative solutions}
\label{sec::approx}

Instead of working with the inverse metric methods discussed in appendix \ref{sec::inverse} to bring the equations into a form that can be treated numerically with an Euler-like algorithm, we present in this section an iterative approach that can be used to obtain approximate solutions to the equations of motion for specific initial and final conditions. Note that for this, it will often be beneficial to rewrite the Lagrangian as  
\begin{align}
\mathrm{L}_{sq}=\iint d\sigma 
d\kappa\ \Pi(\sigma-\kappa)\   
\frac{\dot{f}(\tcs)}{f'(\tcs)} \frac{\dot{f}(\tck)}{f'(\tck)}
=\int d\kappa\ \Pi(\kappa)\int d\sigma\ 
\frac{\dot{f}(\tcs)}{f'(\tcs)} \frac{\dot{f}(\tau,\kappa+\sigma)}{f'(\tau,\kappa+\sigma)}. 
\label{innerintegral}
\end{align}
This identity is of course mathematically trivial, but very helpful in practice when the $\int d\sigma$-integral can be evaluated analytically. Note that for $n\in\mathbb{N}$, \cite{zwillinger2014table}
\begin{align}
&\int_0^{2\pi}d\kappa \log\left(\sin\left(\frac{\kappa}{2}\right)^2\right)\cos(n\kappa)=-\frac{2\pi}{n},
\\
&\int_0^{2\pi}d\kappa \log\left(\sin\left(\frac{\kappa}{2}\right)^2\right)\sin(n\kappa)=0,
\end{align}
so the Lagrangian can then also be evaluated analytically when one knows the Fourier expansion of the $\int d\sigma$-integral. 

Let us assume we want to calculate the geodesic circuit from $f_0(\sigma)=\sigma$ at $\tau=0$ to some  target map,
\begin{equation}
\label{eq.defg}
f_1(\sigma)=\sigma+\varepsilon \, g(\sigma)    
\end{equation} 
at $\tau=1$ with $\varepsilon\ll1$.\footnote{This small expansion parameter $\varepsilon$ should not be confused with the Lie-algebra element $\epsilon(\tcs)$ defined in \eqref{epsilonf}.} We know that to lowest order in $\varepsilon$, the geodesic circuit connecting these two maps will just be the linear interpolation between them, and for higher order corrections in $\varepsilon$ we take the ansatz
\begin{align}
\label{eq.fnF}
f(\tcs)=\sigma+\tau\, \varepsilon \, g(\sigma)+\varepsilon^2\, f^{(2)}(\tcs)+\varepsilon^3 f^{(3)}(\tcs)+... \,.
\end{align}
While the equations of motion~\eqref{EOMf2} at each order in the perturbative expansion in $\varepsilon$ are linear in the sought-after function, they are still IDEs and, therefore, are hard to solve analytically. This can be leveraged when working in Fourier space. To this end, we represent $f^{(n)}(\tcs)$'s from~\eqref{eq.fnF} as a Fourier series
\begin{align}
f^{(n)}(\tcs)=b_{n,0}(\tau)+\sum_{j\in\mathbb{N}}b_{n,j}(\tau)\cos(j\,\sigma)+\sum_{j\in\mathbb{N}}a_{n,j}(\tau)\sin(j\,\sigma).
\end{align}  
Instead of one integro-differential equation of motion for $f^{(n)}(\tcs)$, we obtain an infinite number of coupled ordinary differential equations (ODEs) for the modes $b_{n,0}(\tau)$, $b_{n,j}(\tau)$ and $a_{n,j}(\tau)$. The term $b_{n,0}(\tau)$ is related to the gauge redundancy discussed in section \ref{sec::U1gauge} and can be set equal to zero at every order. The other modes should be solved for subject to the conditions
\begin{equation}
\label{eq.abconditions}
a_{n,j}(0)=a_{n,j}(1)=b_{n,j}(0)=b_{n,j}(1)=0.
\end{equation}


The important realisation is that this problem simplifies greatly when $g(\sigma)$ from~\eqref{eq.defg} is proportional to one single Fourier-mode: $g(\sigma)=\sin(m\sigma+\delta)$ with some integer $m$, where for simplicity we set $\delta=0$. Then, in evaluating the inner $\int d\sigma$-integral in~\eqref{innerintegral} we can make use of identities such as
\begin{align}
\int_0^{2\pi}d\sigma \cos(m\sigma)\cos(n(\sigma+\kappa))=\delta_{m,n}\pi\cos(n\kappa)\text{  for  }m,n\in\mathbb{N},
\end{align}
which allow at each order to obtain an analytical expression for the Lagrangian in terms of the modes $b_{n,j}(\tau)$, $a_{n,j}(\tau)$, and interestingly, high Fourier modes always decouple. To this end, for each order $\varepsilon^n$ of the Lagrangian, and $g(\sigma)=\sin(m\sigma)$ with $m,n\in\mathbb{N}$, there will always be some integer number $M(m,n)$ such that the part of the Lagrangian dependent on these higher modes reads
\begin{align}
\mathrm{L}_{sq}=...+\sum_{j\in\mathbb{N}, j\geq M(m,n)} d_{a,n,j}\ \dot{a}_{n,j}(\tau)^2+ d_{b,n,j}\ \dot{b}_{n,j}(\tau)^2.
\end{align}
This leads to trivial equations $\ddot{a}_{n,j}=0=\ddot{b}_{n,j}$, which, subject to the conditions~\eqref{eq.abconditions}, simply imply $a_{n,j}=0=b_{n,j}$ for sufficiently large $j$. For example, for $g(\sigma)=\sin(\sigma)$, we iteratively find
\begin{align}
f(\tcs)\approx &\sigma +\varepsilon \, \tau \, \sin (\sigma )+\varepsilon^2 \frac{c \, \tau ^2-c \, \tau +20 \, h\, \tau^2-20 \,h\, \tau}{4\, (c+8\, h)} \sin (2 \,\sigma )+...
\label{sinsolution}
\end{align}
with all modes $b_{n,j}=0$. The modes $\sin(m\sigma)$ appearing at order $\varepsilon^n$ in this solution follow the simple pattern shown in table~\ref{tab::modes}. Each time, the function $a_{n,m}(\tau)$ is a simple polynomial in $\tau$ of order $n$.

\begin{table}[h]
	\centering
\begin{tabular}{c|c|c|c|c|c|c|c|c}
	& $m=1$  & $m=2$ & $m=3$ & $m=4$ & $m=5$ & $m=6$ & $m=7$ & $m=8$ \\ 
	\hline 
$n=1$	& X &  &  &  &  &  &  &  \\ 
	\hline 
$n=2$	&  &X  &  &  &  &  &  &  \\ 
	\hline 
$n=3$	& X &  & X &  &  &  &  &  \\ 
	\hline 
$n=4$	&  & X &  & X &  &  &  &  \\ 
	\hline 
$n=5$	& X &  & X &  &X  &  &  &  \\ 
	\hline 
$n=6$	&  & X &  &X  &  &X  &  &  \\ 
	\hline 
$n=7$	&X  &  & X &  & X &  &X  &  \\ 
	\hline 
$n=8$	&  &X  &  &X  &  &X  &  & X \\ 
\end{tabular} 
	\caption{Fourier-modes $\sin(m\,\sigma)$ showing up at order $\varepsilon^n$ in \eqref{sinsolution}.
	}
	\label{tab::modes}
\end{table}

What is now the distance between $f(0,\sigma)=\sigma$ and $f(1,\sigma)=\sigma +\varepsilon\sin (\sigma )$? As the Lagrangian is conserved and the curve is parametrised such that it reaches its destination at $\tau=1$, the square of the distance\footnote{Remember that since section \ref{sec::VIP} we are working with the square of the original Lagrangian, which is more convenient for deriving the equations of motion. However, we have left ambiguous whether the actual value of the complexity should be calculated from this squared Lagrangian, or the original Lagrangian including the square root \eqref{complexityf1}.  See \cite{Bueno:2019ajd} for a discussion of this issue. Having the results of \cite{Flory:2018akz,Belin:2018bpg} for the complexity=volume proposal in mind, the squared Lagrangian seems more relevant for the holographic duality. } is just equal to the value of the Lagrangian $\mathrm{L}_{sq}$ up to a factor of $2\pi$. We find:
\begin{align}
\mathrm{L}_{sq}
=2 \pi ^2 \, \mathit{h} \,\varepsilon ^2+\frac{\pi ^2\, \left(3 \,\mathit{c}^2+56\, \mathit{c}\, \mathit{h}+112 \,\mathit{h}^2\right)}{96\, (\mathit{c}+8 \,\mathit{h})} \varepsilon ^4+...
\label{epsilonseries}
\end{align}
As expected, for small $\varepsilon$ the distance $\sqrt{\mathrm{L}_{sq}}$ will be linear in $\varepsilon$. Note that this order in $\varepsilon$ vanishes for $h=0$, as in this case the initial tangentvector $\dot{f}(0,\sigma)\sim\sin(\sigma)$ of the geodesic circuit for $\varepsilon\rightarrow 0$ is a null-vector, see section~\ref{sec::deg}. Note that for $\varepsilon=1$, $f'(1,\sigma)=1 +\cos (\sigma )$ can be zero, and such maps with $f'(\sigma)= 0$ for some $\sigma$ are not proper group elements as they would not be invertible one-to-one maps of the circle to itself. Given that such maps would also lead to a divergence in $\frac{\dot{f}}{f'}$ and hence the Lagrangian, it seems reasonable that they might form some kind of (asymptotic) boundary of the space of allowed maps, presumably at infinite distance\footnote{While the distance in Hilbert space is an angle between the two states and lies between $0$ and $\frac{\pi}{2}$, this is not the case for the Fubini-Study complexity in most of the cases. For example, the Fubini-Study complexity between two groundstates of harmonic oscillators with frequencies $\omega_{1}$ and $\omega_{2}$ based on Gaussian transformations is proportional to $\left|\log{\frac{\omega_{1}}{\omega_{2}}}\right|$ and can be arbitrarily large~\cite{Chapman:2017rqy}. The ``simple'' rotation between such two states is realized via a non-Gaussian transformation. We believe an analogous statement applies in the case presented here and the possibility of diverging Fubini-Study distance is simply a feature of our construction.}. This might be envisioned to be qualitatively similar to the finite-dimensional hyperbolic disc and we will return to this discussion in sections~\ref{sec::triangle} and~\ref{sec::EA}. 
We would hence expect that the series in \eqref{epsilonseries} has a finite convergence radius $\varepsilon<1$ and describes an analytic function with a pole at $\varepsilon=1$. For this reason we should not trust results obtained with our iterative procedure anywhere near $\varepsilon=1$. We will discuss this further in section \ref{sec::EA}, and will study the limit $h/c\rightarrow\infty$ in some more detail in appendix \ref{sec::effingprimes}. 

One aspect, which would be interesting to understand better, are possible shortcuts in circuits leading to~\eqref{epsilonseries} related to an exchange of dominance (i.e. minimal length) between competing geodesics. Such shortcuts were discussed in a related context in~\cite{Balasubramanian:2019wgd}.

\section{Properties of the Fubini-Study geometry}
\label{sec::geodesics}

\subsection{Geodesic triangles}
\label{sec::triangle}

\begin{figure}[htbp]
	\centering
	\def\svgwidth{0.76\columnwidth}
\executeiffilenewer{Both2.svg}{Both2.pdf}%
{inkscape -z -D --file=Both2.svg %
--export-pdf=Both2.pdf --export-latex}%
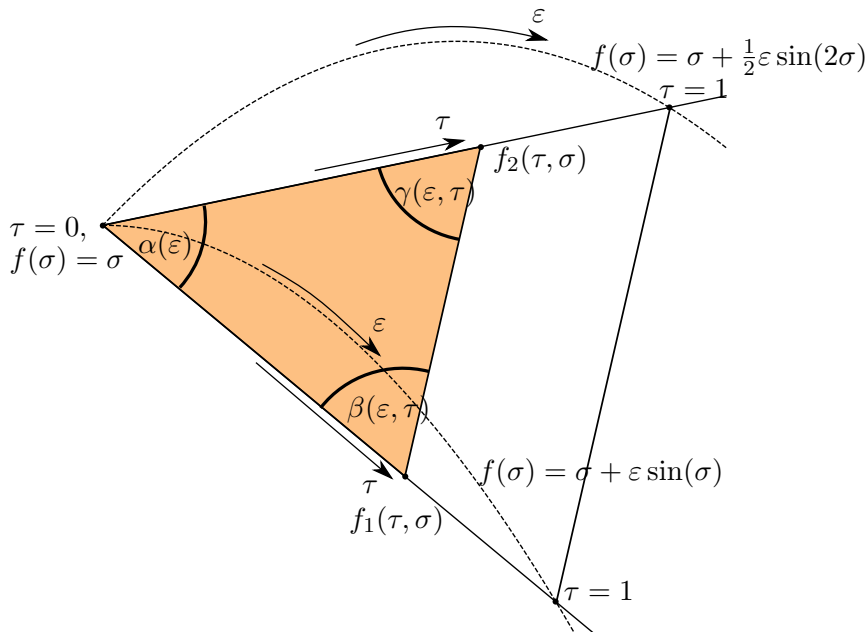%

	\caption{Geodesic triangle in the space of maps. The dashed lines signify the non-geodesic paths $f_1(1,\sigma)=\sigma+\varepsilon\sin(\sigma)$ and $f_2(1,\sigma)=\sigma+\frac{1}{2}\varepsilon\sin(2\sigma)$ parametrised by $\varepsilon$. For small $\varepsilon$, we are able to approximately calculate the geodesics between these endpoints and the identity map $f(\sigma)=\sigma$ via the iterative procedure of section \ref{sec::approx}. These geodesics are parametrized by $\tau$, and we can also calculate the approximate geodesic and distance between the points at $0\leq \tau\leq1$ on these geodesics. All geodesics are shown as straight lines. As a function of $\varepsilon$ and $\tau$, we can now calculate the lengths of the sides of this geodesic triangle as well as the angles in its edges. This allows to probe the sectional curvatures of the space in the respective tangent plane.}
	\label{fig::triangle}
\end{figure}

It would be useful to characterise the geometry of the infinite dimensional space that we are working with by calculating curvature invariants. Unfortunately, as the metric is degenerate, there is no inverse metric and hence Christoffel-symbols, Riemann-tensor and Ricci-scalar cannot be defined. While in the following section we will look at such characterizations of the equivalent metric introduced in appendix~\ref{sec::inverse}, here we focus on directly probing the underlying by studying infinitesimal geodesic triangles and resulting sectional curvatures, see figure~\ref{fig::triangle}. To achieve this, we calculate the geodesic distances between the three points $f(\sigma)=\sigma,\ f_1(\tau,\sigma)$, and $f_2(\tau,\sigma)$. The latter are the points at $0\leq\tau\leq1$ along the geodesics connecting $f_1(0,\sigma)=f_2(0,\sigma)=\sigma$ and $f_1(1,\sigma)=\sigma+\varepsilon\sin(\sigma)$ and $f_2(1,\sigma)=\sigma+\frac{1}{2}\,\varepsilon\sin(2\,\sigma)$. Both geodesics assume~$\varepsilon\ll 1$ and the optimal circuit for $f_{1}(\tau,\sigma)$ is given by~\eqref{sinsolution}.

There are two objects that are related to the sectional curvature which are of interest to us. The first is the sum of the angles of the triangle
\begin{align}
\alpha(\varepsilon)+\beta(\varepsilon,\tau)+\gamma(\varepsilon,\tau)=\pi+\frac{\mathit{h}\,  (13 \,\mathit{h}-4 \,\mathit{c})}{8 (\mathit{c}+3 \,\mathit{h}) \sqrt{\mathit{h}\, (\mathit{c}+8\, \mathit{h})}}\,\tau^2 \varepsilon ^2+...    \,.
\end{align}
As expected, in the limits $\varepsilon\rightarrow0$ or $\tau\rightarrow0$ where the triangle becomes small, the sum of angles goes to $\pi$. This is the correct result on a flat space, and of course locally any curved space looks flat. The next to leading term is interesting, because it indicates that the sum of the angles of the infinitesimal geodesic triangle is $<\pi$ for $h/c<\frac{4}{13}$ (indicative of a negative sectional curvature) and $>\pi$ for $h/c>\frac{4}{13}$ (indicative of a positive sectional curvature). 

Similarly, we can calculate the geodesic distance between $f_1(\tau,\sigma)$ and $f_2(\tau,\sigma)$ as a function of $\tau$:
\begin{align}
d_{1\leftrightarrow2}(\tau,\varepsilon)=\left(\frac{\pi  \sqrt{\mathit{c}+24\, \mathit{h}}}{2 \sqrt{2}} \varepsilon +\mO(\varepsilon^2)\right)\tau+\left(\frac{\pi \, \mathit{h} \, (4 \,\mathit{c}-13\, \mathit{h})}{12 \sqrt{2} (\mathit{c}+3\, \mathit{h}) \sqrt{\mathit{c}+24 \,\mathit{h}}} \varepsilon ^3+\mO(\varepsilon^4)\right)\tau^3  + ...
\end{align}
To leading order, the increase is linear in $\tau$, which again is expected for a flat space. However, the next orders show that the distance between the geodesics will grow faster than linearly for $h/c<\frac{4}{13}$, as again is typical for negatively curved spaces which tend to defocus geodesics. For $h/c>\frac{4}{13}$ the growth of the distance is less than linear, as is typical for positively curved spaces which tend to focus geodesics.   
This demonstrates how we can use our techniques for calculating geodesic distances developed in section \ref{sec::approx} to probe the sectional curvature of the space in a given tangent-plane.    

The reason why the sign of the sectional curvature is interesting is that at the end of section \ref{sec::approx} we formulated our expectation for an analogy between the hyperbolic disk and the space under study here, and according to this analogy we would naively expect to find negative sectional curvatures.  Furthermore, negative sectional curvatures in the dynamics of complex systems are known to be indicative of a strong dependence of geodesics on initial conditions and hence an instability of the geodesic problem~\cite{Arnold2014a,Arnold2014c}, 
and for models of complexity the qualitative importance of negative sectional curvatures has been discussed in detail in~\cite{Brown:2016wib,Brown:2017jil,Brown:2019whu}. We will consequently study sectional curvatures in more detail in the next section, and then discuss the qualitative meaning of our findings in more detail in section~\ref{sec::EA}.

\subsection{Curvature of equivalent metric}
\label{sec::sectional}

As explained in the previous section, we cannot formally compute curvature invariants such as Ricci curvature or sectional curvatures because our metric is degenerate. However, we \textit{can} calculate curvature invariants for the equivalent metric discussed in appendix~\ref{sec::inverse}, and, as we will show below, the values of these invariants will describe the physical behaviour of a well defined set of geodesics in our model, specifically the geodesics investigated in sections \ref{sec::approx} and \ref{sec::triangle}.

In ordinary, finite dimensional differential geometry, the sectional curvature $K(u,v)$ in a tangent-plane spanned by vectors $u^\mu,v^\mu$ is defined as
\begin{align}
K(u,v)=\frac{R_{iklm}u^i u^l v^k v^m}{||u||^2||v||^2-(u\cdot v)^2}.
\label{Kuv}
\end{align}
To make the paper self-contained, we provide below the form of the Riemann tensor
\begin{align}
R_{iklm}=\frac{1}{2}\left(
\frac{\partial^2 g_{im}}{\partial x^k \partial x^l}+\frac{\partial^2 g_{kl}}{\partial x^i \partial x^m}
-\frac{\partial^2 g_{il}}{\partial x^k \partial x^m}-\frac{\partial^2 g_{km}}{\partial x^i \partial x^l}
\right)
+g^{np}\left(\Gamma_{nkl}\Gamma_{pim}-\Gamma_{nkm}\Gamma_{pil}\right)
\label{Riemann}
\end{align}
and the Christoffel-Symbols (of the first kind) are defined in~\eqref{christoffel}. As anticipated, \eqref{Riemann} requires the use of the inverse metric, and can hence not be defined for a degenerate metric, however, it can be easily defined for the invertible equivalent metric \eqref{eqmetriclambda} defined in appendix~\ref{sec::inverse}.

For this, we first note that using the rules of geodesic analogy \eqref{geoan1}-\eqref{geoan3}, 
the equations of motion \eqref{EOMf2} can indeed be phrased in a form equivalent to \eqref{degenerategeodesics} by writing
\begin{align}
\frac{\partial g_{\sigma\alpha}}{\partial x^\beta}\rightarrow \frac{\delta}{\delta f(\beta)}\frac{\Pi(\sigma-\alpha)}{f'(\sigma)f'(\alpha)}=-\Pi(\sigma-\alpha)\left(\frac{\delta'(\sigma-\beta)}{f'(\sigma)^2f'(\alpha)}+\frac{\delta'(\alpha-\beta)}{f'(\sigma)f'(\alpha)^2}
\right).
\end{align}
Similarly, in \eqref{Riemann} we make the replacement
\begin{align}
\frac{\partial^2 g_{\sigma\alpha}}{\partial x^\gamma\partial x^\beta}\rightarrow 
&\frac{\delta}{\delta f(\gamma)}\left(\frac{\delta}{\delta f(\beta)}\frac{\Pi(\sigma-\alpha)}{f'(\sigma)f'(\alpha)}\right)
\\
\nonumber
&=2\delta'(\sigma-\gamma)\delta'(\sigma-\beta)\frac{\Pi(\sigma-\alpha)}{f'(\sigma)^3f'(\alpha)}+\delta'(\alpha-\gamma)\delta'(\sigma-\beta)\frac{\Pi(\sigma-\alpha)}{f'(\sigma)^2f'(\alpha)^2}
\nonumber
\\
&+2\delta'(\alpha-\gamma)\delta'(\alpha-\beta)\frac{\Pi(\sigma-\alpha)}{f'(\sigma)f'(\alpha)^3}+\delta'(\sigma-\gamma)\delta'(\alpha-\beta)\frac{\Pi(\sigma-\alpha)}{f'(\sigma)^2f'(\alpha)^2}.
\nonumber
\end{align}
As said before, in order to evaluate \eqref{Riemann} we need to be able to define an inverse metric. At this point we note that the iterative solutions calculated as in section \ref{sec::approx}  for the sides of a geodesic triangle like the one in figure \ref{fig::triangle} turn out to satisfy condition \eqref{condition} for generic values of $c,h>0$ as long as the corners of the triangle are located at points
$f(\sigma)=0,f_1(1,\sigma)=\sigma+\frac{\varepsilon}{m}\sin(m\,\sigma),f_2(1,\sigma)=\sigma+\frac{\varepsilon}{n}\sin(n\,\sigma)$, $\ n\neq m,\ n,m\in\mathbb{N}$. This means that when studying the geometry of these geodesics we can replace the two-point function by its equivalent \eqref{eqmetriclambda}, without changing the relevant sectional curvatures. The Fourier-components of the inverse of the equivalent metric are then defined by equation \eqref{amalg} for every $m\in\mathbb{Z}$. We then use the geodesic analogy \eqref{geoan1}-\eqref{geoan3} to justify the replacement
$
g^{np}\,\Gamma_{nkl}\,\Gamma_{pim}\rightarrow \iint dn\, dp\, \amalg(n-p)\,f'(\tau,n)\,f'(\tau,p)\,\Gamma(n,k,l)\,\Gamma(p,i,m)
$.

Hence we are able to calculate the extrinsic curvatures \eqref{Kuv} at the location $f(\sigma)=\sigma$ ($\Rightarrow f'(\sigma)=1$) for tangent-vectors of the form $u=\sin(m\,\sigma),v=\sin(n\,\sigma), m\neq n, m,n\in\mathbb{N}$. For this we first note that $u\cdot v=0$ and the norm $||v||^2$ was already given in \eqref{normv}.
The sectional curvatures can then be evaluated in a rather tedious computation, and we obtain the result
\begin{align}
K(u,v)=\frac{3}{\pi ^2\, (m+n)}\left(\frac{(2 \,m+n) (m+2\, n)}{24\, h+c \,(m+n-1) \,(m+n+1)}-\frac{(2\, m-n) \,(m+n)^2}{m \,\left(24\, h+c\, \left(m^2-1\right)\right)}\right)\text{ for $m>n$.} 
\label{KuvValue}
\end{align}
This is one of our most important results. As expected, this result is independent of the constant $\lambda>0$ introduced in appendix~\ref{sec::inverse} to cure the degeneracy of the metric.
Some concrete values for sectional curvatures for $h=0$ and $h=c$ are shown in table \ref{tab::Kuv}. Note that the $h=0$ results are provided under the assumption that we can smoothly take the limit $h\rightarrow 0$ of our results for geodesic circuits and sectional curvatures, despite the enhanced degeneracy of the metric at $h=0$ discussed in sections \ref{sec::deg} and appendix \ref{sec::SL2R}.

\begin{table}[htb]
	\centering
\begin{tabular}{c|ccccc}
$n$ & $m=1$  & $m=2$ & $m=3$ & $m=4$ & $m=5$   \\[5pt] 	
\hline 
$1$	& X & $-\frac{2}{\pi ^2}$ & $-\frac{3}{4 \pi ^2}$ & $-\frac{2}{5 \pi ^2}$ & $-\frac{1}{4 \pi ^2}$  \\[5pt]
$2$	&-0.20 & X & $-\frac{11}{10 \pi ^2}$ & $-\frac{23}{35 \pi ^2}$ & $-\frac{61}{140 \pi ^2}$   \\[5pt]
$3$	&-0.076 & -0.11 & X & $-\frac{43}{56 \pi ^2}$ & $-\frac{461}{840 \pi ^2}$  \\[5pt]
$4$	&-0.041 & -0.067 & -0.078 & X & $-\frac{71}{120 \pi ^2}$ \\[5pt]
$5$	&-0.025 & -0.044 & -0.056 & -0.060 & X  \\[5pt]
\end{tabular}
\  
\begin{tabular}{c|ccccc}
$n$	& $m=1$  & $m=2$ & $m=3$ & $m=4$ & $m=5$    \\[5pt] 	
\hline 
$1$	& X & $\frac{1}{8 \pi ^2}$ & $\frac{5}{104 \pi ^2}$ & $\frac{1}{520 \pi ^2}$ & $-\frac{53}{2360 \pi ^2}$ \\[5pt]
$2$	&0.013 & X & $\frac{3}{40 \pi ^2}$ & $-\frac{11}{767 \pi ^2}$ & $-\frac{2}{35 \pi ^2}$  \\[5pt]
$3$	&0.0049 & 0.0076 & X & $-\frac{5}{273 \pi ^2}$ & $-\frac{97}{1160 \pi ^2}$  \\[5pt]
$4$	&0.00020 & -0.0015 & -0.0019 & X & $-\frac{11}{120 \pi ^2}$  \\[5pt]
$5$	&-0.0023 & -0.0058 & -0.0085 & -0.0093 & X  \\[5pt]
\end{tabular} 
	\caption{Sectional curvatures $K(u,v)$ at $f(\sigma)=\sigma$ for $c=1, h=0$ (top) and $c=h=1$ (bottom) for small values of $m,n$. Note that \eqref{KuvValue} is not symmetric under exchanging $m\leftrightarrow n$ because it was simplified assuming $m>n$ in order to get rid of terms like $\text{sign}(m-n)$, but the values of $K(u,v)$ are of course symmetric. We make use of that symmetry, for the convenience of the reader, to display both exact values above the diagonal and approximate decimal values below the diagonal. While in the $h=0$ case all sectional curvatures are negative, for large enough values $h$ a finite number of the sectional curvatures will be positive. Only in the limit $h/c\rightarrow \infty$ will all the curvatures become positive, see also table \ref{tab::hcrit}.}
	\label{tab::Kuv}
\end{table}

One can show that for $h/c>\frac{4}{13}$, a finite number of these sectional curvatures will be positive, while infinitely many ones for large enough $m,n$ will be negative. This seems to indicate that the \textit{generic} curvature felt by the geodesic curves we are investigating will be negative. 
Only in the limit $\frac{h}{c}\rightarrow\infty$ do all sectional curvatures in \eqref{KuvValue} become positive. The critical value of $\frac{h}{c}$ at which the sectional curvature in a given tangent-plane switches sign from negative to positive is given by
 \begin{align}
\left( \frac{h}{c}\right)_{crit}=\frac{2 m^4+6 m^3 n+2 m^2 n^2+2 m^2-2 m n^3+2 m n-n^4+n^2}{24 \left(2 m^2+2 m n+n^2\right)}\text{ for $m>n$,} 
\label{hcrit}
 \end{align}
with sample values for small $m,n$ provided in table~\ref{tab::hcrit}. Clearly, for $m=2,n=1,$ we recover the value of $\frac{4}{13}$ that played a prominent role in section \ref{sec::triangle}, which is an important consistency check between the two approaches.

\begin{table}[htb]
	\centering
\begin{tabular}{c|cccccc}
	& $m=1$  & $m=2$ & $m=3$ & $m=4$ & $m=5$ & $m=6$   \\[5pt] 	
\hline 
$n=1$	& X & $\frac{4}{13}$ & $\frac{3}{5}$ & $\frac{40}{41}$ & $\frac{175}{122}$ &$ \frac{168}{85}$ \\[5pt]
$n=2$	&0.308 & X & $\frac{11}{17}$ & $\frac{115}{104}$ & $\frac{61}{37}$ & $\frac{91}{40}$ \\[5pt]
$n=3$	&0.600 & 0.647 &X & $\frac{43}{39}$ & $\frac{461}{267}$ & $\frac{95}{39}$ \\[5pt]
$n=4$	&0.976 & 1.11 & 1.10 & X & $\frac{355}{212}$ & $\frac{335}{136}$ \\[5pt]
$n=5$	&1.43 & 1.65 & 1.73 & 1.67 &X & $\frac{371}{157}$ \\[5pt]
$n=6$	&1.98 & 2.27 & 2.44 & 2.46 & 2.36 & X \\[5pt]
\end{tabular} 
	\caption{Values of $h/c$ at which the sectional curvatures $K(u,v)$ turn from negative to positive, for small values of $m,n$. This is calculated from equation \eqref{hcrit} for $m>n$, and as in table \ref{tab::Kuv} we display both exact and approximate values. }
	\label{tab::hcrit}
\end{table}

We can now study additional tangent-planes at $f(\sigma)=\sigma$. For tangent vectors of the form $u=\sin(m\,\sigma),v=\cos(n\,\sigma), m\neq n, m,n\in\mathbb{N}$, we still have $u\cdot v=0$ and the norms given by  \eqref{normv}. In fact, for $m>n$ the sectional curvatures are given by \eqref{KuvValue} again, while for $m<n$ we would get the same expression \eqref{KuvValue} with $m\leftrightarrow n$ interchanged. 
For $u=\sin(m\,\sigma),v=\cos(n\,\sigma)$ we can also study the case $m=n, n\in\mathbb{N}$. We still have $v\cdot u=0$ and 
 \begin{align}
K(u,v)= \frac{3 \, n}{\pi ^2} \left(\frac{4}{\mathit{c}-24 \,\mathit{h}-\mathit{c}\, n^2}-\frac{9}{\mathit{c}-24 \,\mathit{h}-4 \,\mathit{c} \,n^2}\right)-\frac{432 \,\lambda }{\left(24 \,\pi \, \mathit{h}+\pi \, \mathit{c} \,\left(n^2-1\right)\right)^2}.
\label{KuvValue2}
\end{align}
Note that in contrast to \eqref{KuvValue}, this curvature now depends explicitly on the parameter $\lambda$ introduced in appendix~\ref{sec::inverse}, however, the limit $\lambda\rightarrow0$ can be smoothly taken. For $h=0$ and $n=1$, this curvature diverges because $||v||^2=0$, in contrast to \eqref{KuvValue} where even in this limit we obtain a finite result as factors from the nominator and denominator cancelled in the derivation. Similar to the above results we find that for $h/c\ll1$, all the (finite) curvatures \eqref{KuvValue2} will be negative, and as $h/c$ increases for given $n$ will turn positive at the critical values 
 \begin{align}
\left( \frac{h}{c}\right)_{crit}=\frac{1}{120} \left(5+7\, n^2\right),
\label{hcrit2}
 \end{align}
where we have already taken the limit $\lambda\rightarrow0$. This is again consistent with our earlier finding that for fixed finite $h/c\geq0$, at most a finite number\footnote{Of course, in saying this we only consider sectional curvatures for tangent planes spanned by orthogonal tangent vectors of the type $u=\sin(m\sigma)$ or $v=\cos(n\sigma)$, $m,n\in\mathbb{N}$, such that there is a countable infinity of such tangent planes.} of the sectional curvatures will be positive, while the remaining infinite number of sectional curvatures will be negative, so that the \textit{generic} curvature of the space is negative unless the limit $h/c\rightarrow+\infty$ is taken. As discussed in the previous section, the sign of the sectional curvatures is qualitatively important for the physical interpretation of our distance measure as a measure of complexity. We will discuss this issue in more detail in the next section, where we compare the geometry investigated in this paper with the one underlying the Korteweg-de Vries (KdV) and other Euler-Arnold type equations.


\section{Comparison to KdV and other Euler-Arnold equations}
\label{sec::EA}

\subsection{Euler-Arnold equations as wave equations}

So far, we have defined a form of geodesic motion on the infinite-dimensional Virasoro group-manifold via the functional \eqref{complexity} and the equations of motion \eqref{EOMf2}. As we explained, these equations can be readily recognized by any general relativist as a generalisation of geodesic motion to infinite dimensions via the geodesic analogy of section \ref{sec::ga}. Especially the fact that summation over discrete indices is replaced by integration over a continuous variable \eqref{geoan1} is important as it leads to the integro-differential nature of the equation \eqref{EOMf2}. 
The concept of geodesic motion on group-manifolds including the Virasoro-group is not new and has been studied in detail to date, see \cite{khesin2008geometry,LeeThesis,Vizman2008,arnold2013vladimir} for overviews. Such equations are often referred to as Euler-Arnold equations, due to the seminal contributions of Arnold \cite{arnold2013vladimir}, who amongst other achievements showed that the Euler equations of fluid dynamics can be phrased as a geodesic motion on a diffeomorphism group \cite{Arnold2014b}.

The usual procedure herein is to define an inner product on the Lie-algebra of the group, which is extended to the tangent-space at every group-element by requiring right- or left-invariance. For example, on the Virasoro-group with $(v(\tcs)\partial_\sigma\,,b^{(v)}(\tau))$ being a two-component object consisting of the Lie-algebra element $v(\tcs)\partial_\sigma$ of the group of diffeomorphisms 
and the real-valued function function $b^{(v)}(\tau)$ associated with the central extension~\cite{Oblak:2016eij}, the inner product 
\begin{align}
 \left\langle (v(\tcs)\partial_\sigma\,,b^{(v)}(\tau)),\,(w(\tcs)\partial_\sigma\,,b^{(w)}(\tau))\right\rangle \equiv \int \left(\alpha\, v\,w+\beta\, v'\,w'\right)\, d\sigma +b^{(v)}\cdot b^{(w)}
\label{virasoroproduct}
 \end{align}
leads to the equations of motion
  \begin{align}
\alpha \left(\dot{v}+3v\,v'\right)-\beta \left(\dot{v}''+2v'v''+v\,v'''\right)-b^{(v)}v'''=0,
\label{KdV}
\\
\dot{b}^{(v)}=0.
\label{dot}
\end{align}
A number of famous integrable equations follow from this, such as the Camassa-Holm equation ($\alpha=\beta=1$), the Hunter-Saxton equation ($\alpha=0, \beta=1$), the Korteweg-de Vries equation ($\alpha=1,\beta=0$) and the inviscid Burgers equation ($\alpha=1,\beta=0,b^{(v)}(\tau)=0$). 

All the aforementioned equations have the well-known physical interpretation of describing wave propagation and we will first briefly review this point of view following~\cite{Oblak:2020jek} and in section~\ref{sec::EArelationtocomplexity} elucidate their possible interpretation in terms of complexity. The use of Euler-Arnold equations in the latter context was proposed in~\cite{Caputa:2018kdj} and were subsequently explored in~\cite{Balasubramanian:2019wgd,Erdmenger:2020sup} and our previous paper~\cite{Flory:2020eot}. Consider again the KdV equation~\eqref{KdV} ($\alpha=1,\beta=0$) and set $b^{(v)}=\frac{c}{12}\neq0$, consistent with \eqref{dot}. The $2\pi$-periodic function $v(\tcs)$ can be interpreted as the \textit{wave profile} of a shallow water wave. Of course, we know that waves propagating in media may transport energy, but they do not always have to transport particles. The fluid-elements whose collective movement forms the wave may not have travelled at all from their initial position after a wave has passed. Following \cite{Oblak:2020jek}, the time-dependent position $x(\tau)$ of a fluid element shuffled around by the passing wave with wave profile $v$ (which solves the KdV equation) is given by 
\begin{align}
   \dot{x}(\tau)=v(\tau,x(\tau)). 
   \label{oblak1}
\end{align}
Some interesting general properties of solutions $x(\tau)$ to this equation were studied in \cite{Oblak:2020jek}. 
Comparing with our formulation and notation of circuits describing conformal transformations of section \ref{sec::conformal} and based on~\cite{Caputa:2018kdj,Erdmenger:2020sup,Flory:2020eot}, we can identify $v$ with $\epsilon$ (as expected, they are the Lie-algebra elements), $x$ with $f$ (they are positions on the circle), and equation \eqref{oblak1} with \eqref{epsilonf}. 

One important phenomenon that has been studied extensively in such wave equations is \textit{wave breaking}. For example in the case of the HS equation, it was shown in \cite{Lenells07} that the underlying geometry maps the group-manifold to an open subset of a sphere. At the end of section \ref{sec::approx}, we discussed that as we deal only with \textit{invertible} maps $f(\tcs)$ of the circle to itself, the maps for which the derivative $f'(\tcs)$ can either vanish or diverge form a boundary of this set. The result of \cite{Lenells07} implies that under the geometry implied by the HS equation, this set is geodesically incomplete. Geodesics can leave the space of invertible maps on the circle in finite affine parameter $\tau$, corresponding to wave profiles that break in finite time. 

\subsection{Relation to our framework \label{sec::EArelationtocomplexity}}

We have seen that the $v$ and $w$ in \eqref{virasoroproduct} are elements of the Virasoro-algebra, i.e.~correspond to $\epsilon(\tcs)$~\eqref{epsilonf} in our notation, and we ignored the component $b^{(v)}(\tau)$ related to the central extension since it gives rise to a phase factor in~$U$~\eqref{eq.circuit}. An alternative way to view~$\epsilon$ is
 \begin{align}
 \epsilon(\tau,f(\tcs))=\dot{f}(\tcs)\Rightarrow  \epsilon(\tau,\sigma)=-\frac{\dot{F}(\tcs)}{F'(\tcs)},
 \label{epsilon}
 \end{align}
 where $F$ is the inverse of $f$, i.e.
 \begin{equation}
     F(\tau,f(\tcs))=\sigma.
 \end{equation} 
 For example, to derive the Burgers equation directly from \eqref{virasoroproduct} with $\alpha=1,\,\beta=0$, we first note that in terms of the circuit $f$, we have $\int \epsilon(\tcs)^2 \,d\sigma = \int \dot{f}(\tau,\rho)^2\,f(\tau,\rho)\,d\rho$ for the kinetic term of the geodesics. The equation of motion following from this after variation with respect to $f$ reads
 \begin{align}
 0=\ddot{f}f'+2\dot{f}\dot{f}'\Rightarrow 0=\dot{\epsilon}+3\epsilon\epsilon',
 \end{align}
 where in the last step we recover the Burgers equation in its usual form by rephrasing the equation in terms of $\epsilon$ instead of $f$. 

In order to connect complexity-born functional~\eqref{complexityf2} with inner products of the form \eqref{virasoroproduct}, we recognize the ratio~$\frac{\dot{f}(\tcs)}{f'(\tcs)}$ appearing there as an analogue of $\epsilon$ for the inverse circuit
 \begin{align}
\hat{ \epsilon}(\tau,F(\tcs))=\dot{F}(\tcs)\Rightarrow  \hat{\epsilon}(\tau,\sigma)=-\frac{\dot{f}(\tcs)}{f'(\tcs)}.
\label{hatepsilon}
 \end{align}
It is then clear that \eqref{complexityf2} can be phrased as only being dependent on $\hat{\epsilon}$. 

In order to understand how our approach fits together with the Euler-Arnold equations discussed above, let us for a moment ignore the difference between $\epsilon$ and $\hat{\epsilon}$. As discussed, the starting point for Euler-Arnold type equations is an inner product \eqref{virasoroproduct} on the Virasoro-algebra, which has the general form
\begin{align}
 \left\langle (v,b^{(v)}),(w,b^{(w)})\right\rangle_{Vir} \equiv 
 \left\langle v,w\right\rangle_{Diff^{+}}  
 +
  \left\langle b^{(v)},b^{(w)}\right\rangle_{\mathbb{R}}.
  \label{generalproduct}
\end{align}
The norm of a curves tangent-vector is then interpreted as a kinetic term from which geodesic equations for an optimal circuit can be derived. 
As we see, the Virasoro inner product \eqref{generalproduct} is a sum of a product on the algebra of diffeomorphisms of the circle and a product for the real-valued functions $b^{(v)}$ that is a consequence of the central extension. Continuing to ignore the difference between $\epsilon$ and $\hat{\epsilon}$, we can now interpret \eqref{complexityf2} as arising from an expression
\begin{align}
     \left\langle v,w\right\rangle_{Diff^{+}} =\iint d\sigma\, 
	d\kappa\,  \Pi(\sigma-\kappa)  \,
	v(\sigma)\,w(\kappa) .
	\label{generalus}
\end{align}
While one might claim that the Fubini-Study complexity only defines geodesic motion on the group of diffeomorphisms and not the Virasoro group with its central extension, we would rather take the viewpoint that \eqref{generalus} defines an inner product of the form \eqref{generalproduct} on the entire Virasoro-algebra, which however is \textit{degenerate} because we implicitly set $\left\langle b^{(v)},b^{(w)}\right\rangle_{\mathbb{R}}=0$. 
Of course, starting from section \ref{sec::deg} the degeneracy of our metric has been discussed in various places in this paper, and we have come to appreciate it as a very desirable feature of our construction, in line with our generic expectation that operations which only change a state by a complex phase should not incur any cost. 


We now also understand that the inner product \eqref{virasoroproduct} falls neatly into our general framework \eqref{generalus} when choosing
\begin{align}
\Pi(\sigma-\kappa)=\alpha\,\delta(\sigma-\kappa)+\beta\,\delta''(\sigma-\kappa).
\label{ultralocal}
\end{align}
In fact, we can now make educated guesses about which setups might give rise to complexity-equations of this type, and which of the three integrable equations discussed above, KdV, CH and HS, would be the most realistic models of complexity. Firstly, we would expect such equations to describe Fubini-Study state complexity in a hypothetical setup where the two-point function evaluated in $|\mR\rangle$ takes the ultralocal form \eqref{ultralocal}. Secondly, given the importance of differential regularisation (section \ref{sec::diffreg}), metric degeneracy (section \ref{sec::deg}) and gauge invariance (section \ref{sec::U1gauge}) for the absence of cost assignment for complex phases, we recognise that it is the HS equation, $\alpha=0, \, \beta=1$ in \eqref{KdV}, which should yield the most plausible model of complexity\footnote{On a formal level, the degeneracy of the metric on the Virasoro-algebra means that it might be as well defined as a non-degenerate metric on the homogeneous space $Vir/S^1$  \cite{khesin2008geometry}. This operation of taking the group manifold modulo rotations corresponds to fixing the $U(1)$-gauge symmetry of section \ref{sec::U1gauge}.}.
When it comes to the ease of solving, the HS equation lies somewhere in between the KdV equation studied in \cite{Caputa:2018kdj,Erdmenger:2020sup}, which however assigns non-zero complexity to complex phases, and our Fubini-Study result~\eqref{EOMf2}, which are considerably more complicated due to their partial integro-differential nature. 

One interesting physical quantity by which such equations and their solutions can be qualitatively characterized and compared are sectional curvatures. As shown in section \ref{sec::sectional}, the \textit{generic} curvature felt by the geodesic curves solving the Fubini-Study IDE~\eqref{EOMf2} appears to be negative unless $h/c\rightarrow\infty$. This indicates a strong dependence of geodesics on initial conditions \cite{Arnold2014a,Arnold2014c} which is qualitatively expected for models of holographic complexity \cite{Brown:2016wib,Brown:2017jil,Brown:2019whu}, as it is an early-time signal for 
chaos\footnote{A late-time signal for chaos is \textit{ergodicity}. We leave the question of ergodicity in our model for further study, but qualitatively we do not expect it to be present. The reason for this is that we expect circuits in our model to run off towards the (asymptotic) boundary of the space of maps. This is in contrast to the toy-model of \cite{Brown:2016wib}, which is constructed by starting with geodesic motion on a hyperbolic disk and excising the asymptotic boundary, replacing it with a topologically non-trivial "sewing together" of patches of the excision surface. The topological sewing sends geodesics that would go towards the asymptotic boundary back inwards.}. However, we should remind the reader that our equation~\eqref{EOMf2} is completely universal and negative sectional curvature arise for free CFTs$_{1+1}$ as well as for holographic CFTs$_{1+1}$. It is also known that the metrics associated with the KdV and CH equations give rise to sectional curvatures of non-definite sign \cite{10.2307/2161606,MISIOLEK1998203}, while the geometry of the HS equation corresponding to an open subset of a sphere has consequently constant positive curvature~\cite{Lenells07}. As we discussed above, this leads to a geodesic incompleteness of the Virasoro group under the geometry imposed by the HS equation. This may have a perfectly acceptable and interesting physical interpretation as wave breaking when viewing the HS equation as a wave equation, but from a complexity point of view this phenomenon would be harder to interpret, and may call into question the validity of the HS equation as a model of complexity.  


It would thus be valuable to develop a better understanding of the conditions on a generic $\Pi$ under which equations of the form \eqref{EOMf2} do or do not allow for such wave breaking in finite time $\tau$. This issue might be closely connected with other relevant geometric aspects, such as the existence of conjugate points or cut loci \cite{Brown:2019whu,Balasubramanian:2019wgd}, which in turn will clearly depend on the sectional curvatures and their signs. The existence of cut loci or conjugate points would also pose questions about the iterative procedure defined in section~\ref{sec::geodesics}, for example about when it converges to a \textit{global} minimum of the action, or what its convergence radius is in general. 
Of course such questions have been studied in detail for the KdV, CH, and HS equations in the mathematical literature, but we are not aware of generic results for abstract equations of the form \eqref{EOMf2} with general $\Pi$. Interestingly, however, the generalised Constantin-Lax-Majda (CLM) equation studied in \cite{Okamoto_2008,doi:10.1142/S1402925110000544,BAUER2016478} appears to be related (again, up to interchanging $\epsilon$ and $\hat{\epsilon}$) to the inner product \eqref{diffreg-1} in the limit $c\rightarrow 0$. In fact, the sectional curvatures calculated in Theorem 21 of \cite{BAUER2016478} are in exact agreement, up to an overall rescaling of the metric, with the $c\rightarrow0$ limit of our results \eqref{KuvValue}, \eqref{KuvValue2} (for $\lambda=1$). We consider this to be a non-trivial consistency check of our results.

\subsection{Fubini-Study vs.~operator complexity \label{sec::FSvsO}}

Let us now elaborate on the difference between $\epsilon(\tcs)$ as appearing in~\eqref{epsilon} and $\hat{\epsilon}(\tcs)$ defined in~\eqref{hatepsilon}. Obviously, interchanging $\epsilon$ and $\hat{\epsilon}$ corresponds to interchanging $f(\tcs)$ with its inverse function $F(\tcs)$. In the following we will use capital letters to denote inverses.
Note that if $f_2(\sigma)=g(f_1(\sigma))$, then the inverses of these functions satisfy $F_2(\sigma)=F_1(G(\sigma))$, hence the manifest invariance of $\hat{\epsilon}$ under conformal maps discussed in section \eqref{sec::conformal} means our inner product is left-invariant, while the inner product \eqref{virasoroproduct} formulated in terms of $\epsilon$ is invariant under $F\rightarrow G(F)$ and hence right-invariant. 
Note also that according to our complexity definition, the distance between the identity map $f=\sigma$ and a map $f=f_1$ is identical to the distance between $f=\sigma$ and $f=F_1$. This is easy to show by using invariance under a change of affine parameter $\tau\rightarrow s=1-\tau$ and invariance under conformal transformations \eqref{conformalsymmetry}. We thus consider both left- and right-invariant inner products to be equally admissible on physical grounds.

In our view, the difference between using $\epsilon$ and $\hat{\epsilon}$ corresponds to an important difference in the kind of complexity we are defining. 
The first, which we have assumed in all of the paper before this section, emerges by starting with an energy eigenstate $|h \rangle$ and evaluating the Hilbert space distance traversed by the circuit defined by \eqref{eq.circuit}. Herein, the state with respect to which the cost of a generator is counted is constantly updated along the circuit according to \eqref{psioftau}. This is why equation \eqref{eq.costFS} depends on the expectation value $\left\langle \psi(\tau)| Q^2|\psi(\tau)\right\rangle$ of the generator~$Q$~\eqref{eq.Qcft} under the state $\left|\psi(\tau)\right\rangle$. Only by introducing the operator $\tilde{Q}$ in equations \eqref{tildeQ} can we write this in terms of a two-point function in a fixed state $\hrr$. This is of course an instance of the Fubini-Study complexity introduced in \cite{Chapman:2017rqy}, and as we have seen, this approach naturally gives rise to the expressions \eqref{tildeQ2} and \eqref{complexityf2} which manifestly depend on $\hat{\epsilon}(\tcs)$~\eqref{hatepsilon}. This type of complexity should be naturally understood as a \textit{state complexity}, i.e.~a notion of complexity defined directly on the space of states.

A different type of complexity which would be more faithful to the ideas of unitary complexity~\cite{Nielsen:2005mn1,Nielsen:2006mn2,Nielsen:2007mn3,Jefferson:2017sdb} is instead based on the variance of $\left\langle \mA| Q(\tau)^2|\mA\right\rangle$ in a fixed auxiliary state $|\mA\rangle$~\eqref{eq.costvar}. In this case, the cost of each layer depends only on $\epsilon(\tau,\sigma)$ as the insertions of $T$ would be assigned the same weight at each layer. Starting from~\eqref{eq.Qcft}, this would lead to a cost function similar to~\eqref{complexityf2}, but in which $\hat{\epsilon}$ is replaced by $\epsilon$. This type of complexity should be understood as an \textit{operator complexity}, defined on the space of operators independently of the states on which the operators will act, apart from the choice of one fixed auxiliary state $|\mA\rangle$. One natural question to consider is what such definition of cost really counts, i.e. what operators diagonalize\footnote{A general Euclidean norm, with a slight abuse of terminology including possibly also null directions, leads to the cost function $\int_{0}^{1} d\tau \sqrt{\sum_{I,J} \eta_{IJ} Y^{I}Y^{J}}$ with $\eta_{IJ}$ being a constant non-negative definite matrix -- the tanget space metric. Diagonalizing the tangent space metric is equivalent to picking particular linear combinations of operators ${\cal O}_{I}$, denoted by ${\cal O}_{\tilde{I}}$, as fundamental gates. The eigenvalues~$\lambda_{\tilde{I}} \geq 0$ of the tangent space metric correspond then to assigning independent infinitesimal costs to applications of operators from this new basis according to $d\tau \sqrt{\sum_{\tilde{I}} \lambda_{\tilde{I}} \, (Y^{\tilde{I}})^2}$. Therefore, the question raised in the text concerns finding ${\cal O}_{\tilde{I}}$ such that $\Pi(\sigma - \kappa)$ is diagonal, where $\sigma$ and $\kappa$ stand for indices $I$ and $J$.} the tangent space metric~$\Pi(\sigma - \kappa)$. Based on formulas~\eqref{normv} and the study in section~\ref{sec::sectional}, the independent operators whose cost we account for via~\eqref{eq.costvar} are simply Hermitian combinations of $L_{n}$ and $L_{-n}$ Virasoro generators~\eqref{VirasoroAlgebra}.

Note also that the connection with expectation values in some states is not needed for operator complexity, as can be apparent from our discussion in section~\ref{sec::Intro}. If one considers, for example, equation~\eqref{ultralocal} with $\alpha = 1, \, \beta = 0$ corresponding to the Burgers equation, then following section~\ref{sec::Intro} one can interpret it as counting $T(\sigma)$ insertions at each layer of the circuit with the most symmetric Euclidean norm.

In conclusion, we are led to believe that formulating the kinetic term in terms of $\hat{\epsilon}$ corresponds to the definition of a left-invariant state-dependent Fubini-Study complexity, while using $\epsilon$ instead, as usually done for the KdV, CH and HS equations, corresponds to a right-invariant notion of unitary complexity.


\section{Summary and outlook}
\label{sec::conc}

Our article builds on~\cite{Caputa:2018kdj,Flory:2018akz,Belin:2018bpg,Camargo:2019isp,Erdmenger:2020sup} and provides a comprehensive treatment of complexity associated with constructing Virasoro group elements and states obtained by their action on the vacuum or other energy eigenstates in CFTs$_{1+1}$. A compressed version of some of our findings appeared earlier in~\cite{Flory:2020eot}. The main motivation to address this particular problem is shared with~\cite{Caputa:2017urj,Caputa:2017yrh,Czech:2017ryf,Magan:2018nmu,Caputa:2018kdj,Belin:2018bpg,Flory:2018akz,Flory:2019kah,Camargo:2019isp,Chen:2020nlj,Erdmenger:2020sup} and stems from identifying the group of local conformal transformations as a territory where holographic complexity proposals~\cite{Susskind:2014rva,Stanford:2014jda,Couch:2016exn,Brown:2015bva,Brown:2015lvg} stand a chance of being matched by a bona fide QFT calculation based on~\cite{Chapman:2017rqy,Jefferson:2017sdb}. An indication this intuition is correct comes from~\cite{Belin:2018bpg} and uses the definition of state complexity from~\cite{Chapman:2017rqy}, which also plays the central role in our work, combined with the results of gravity calculations from~\cite{Flory:2018akz}.

The setup we adopted, see section~\ref{sec:Conftrans}, follows~\cite{Caputa:2018kdj,Erdmenger:2020sup} and is a single copy of the Virasoro group in a CFT$_{1+1}$ on a circle. The key kick-starting impulses behind our work are twofold and lie in recognizing that a good definition of complexity should be insensitive to complex phases appearing in unitary transformations, as well as lead to a well-posed variational problem for the cost function with two arbitrary group elements defining initial and final conditions for the circuit. Building on observations and efforts made in~\cite{Caputa:2018kdj,Erdmenger:2020sup} we reconsider here the Fubini-Study complexity of states~\cite{Chapman:2017rqy} as a starting point for a well-defined complexity notion assigned to the~Virasoro group and states generated by it. The fact that Fubini-Study complexity is insensitive to complex phases is its built-in feature~\cite{Chapman:2017rqy} and well-posedness of the variational problems comes  from the cost function being quadratic in tangent space velocities~\eqref{complexity}.

As compared to earlier works, we believe that the present article together with our earlier paper~\cite{Flory:2020eot} bring three key advances that we summarize below one by one.

The first one relies on reaching the ability of solving the resulting integro-differential equations of motion and finding the optimal circuits in a class of transformations defined by~\eqref{eq.defg} with $g(\sigma)$ containing only a single Fourier mode and $\varepsilon$ being sufficiently small. This is the subject of section~\ref{sec::affine} and uses preparatory studies done in section \ref{sec::FSmetric}. Being able to solve for optimal circuits in the Fubini-Study state complexity is not only interesting as a technical achievement, or a vehicle to shed light on the underlying geometry as we describe next, but also provides concrete numbers that one can compare against predictions of the holographic complexity proposals. We will comment more on this point when we discuss open directions.

The second advance starts with what we call a geodesic analogy, see section~\ref{sec::ga}, and encapsulates studies of the underlying geometry of states close to the reference state utilizing geodesic triangles known from general relativity, as well as more directly curvature invariants via the notion of sectional curvature, see section~\ref{sec::geodesics}. It is important to note that in the latter case we cannot use directly expressions known to physicists from general relativity, since the underlying metric is degenerate, see section~\ref{sec::deg}. We circumnavigate this issue by utilizing what we call the equivalent metric method, see appendix~\ref{sec::inverse}. One important finding of our study is that the Fubini-Study state complexity leads to generically negative sectional curvatures in the underlying circuit geometry in what we view as the physically interesting regime, i.e.~unless $h/c \rightarrow \infty$. The relation between complexity and negative curvature was discussed earlier in~\cite{Brown:2016wib,Brown:2017jil,Brown:2019whu} and one can perceive our work as a concrete and precise derivation of such a feature across CFTs$_{1+1}$.

The third contribution that we want to highlight is bringing the Euler-Arnold equations discussed in the same context in~\cite{Caputa:2018kdj,Erdmenger:2020sup} and our studies of Fubini-Study metric under single conceptual umbrella, see section~\ref{sec::EA}. Specifically, we are now in a position to predict that, up to issues of left versus right-invariance, well-known Euler-Arnold type equations like Korteweg-de Vries~\cite{Caputa:2018kdj,Erdmenger:2020sup}, Camassa-Holm and Hunter-Saxton will naturally arise from a Fubini-Study ansatz when a combination of reference state and generator set is used such that the two-point function takes the form \eqref{ultralocal}. We conjecture that the relevant reference state would be a state without spatial entanglement. Such states arise in tensor networks as simple states~\cite{Haegeman:2011uy} and are expected to play an important role in holography~\cite{Miyaji:2015fia}. If indeed~\eqref{ultralocal} can be thought of as a two-point function in such a state, then, via~\eqref{eq.costvar}, our work would provide a unified view on a state and operator complexity associated with the Virasoro group.

While the Korteweg-de Vries, Camassa-Holm, and Hunter-Saxton equations are certainly among the most well-known Euler-Arnold type equations, they are by far not the only ones, and as discussed in section \ref{sec::EA} the limit $c\rightarrow0$ of our metric corresponds to the generalised Constantin-Lax-Majda equation studied in~\cite{Okamoto_2008,doi:10.1142/S1402925110000544,BAUER2016478}. Without any doubt, the mathematical literature on Euler-Arnold type equations is a treasure trove of known equations, results and methods, that could be exploited by the community interested in holographic and/or QFT complexity.
In our paper, we have only made some first steps into this programme, and a more formal treatment of our equations might certainly be enlightening, especially in light of previous results such as~\cite{Zumino1988,PhysRevLett.58.535,FANG2002162}.

Regarding further open problems, one direction that certainly deserves further studies is comparison of our results on the Fubini-Study complexity with the predictions of holographic complexity proposals in the setting of AdS$_{1+2}$ gravity~\cite{Flory:2018akz,Belin:2018bpg,Flory:2019kah}. As we already indicated, a match of leading order results for infinitesimal transformations with the complexity = volume proposal was reported in~\cite{Belin:2018bpg} and it would be very interesting to make a comparison also at the level of the terms higher order in~$\varepsilon$. One feature that we want to highlight is that the aforementioned agreement was observed for~\eqref{innerintegral} evaluated on an optimal circuit rather than directly for~\eqref{complexity}.

It should be noted though that the direction outlined above is not meant to be an apple-to-apple comparison, since holographic complexity proposals, assuming they indeed represent some notion of complexity in dual QFTs, are expected to take as their reference state a spatially disentangled one~\cite{Chapman:2017rqy,Jefferson:2017sdb}, whereas our studies concern states whose ultraviolet entanglement is as in the vacuum. It should be noted that recent results~\cite{Belin:2020oib,Kruthoff:2020hsi} on circuits utilizing $T\bar{T}$ deformations~\cite{Smirnov:2016lqw,Cavaglia:2016oda,McGough:2016lol} provide a promising starting point to establish a direct link between AdS volumes and gate counting in holographic QFTs, which would be a quantitative realization of a generalization of the path-integral optimization program~\cite{Caputa:2017urj,Caputa:2017yrh,Czech:2017ryf,Takayanagi:2018pml,Camargo:2019isp} akin to~\cite{Jafari:2019qns,Geng:2019yxo}.

The discussion in the previous paragraph raises another interesting point -- can one realize the Fubini-Study complexity, as we define it, in holography with a natural gravity dual? We believe this is a very interesting point, since one of the drawbacks of our existing formulations of holographic complexity is the absence of explicit control over reference states.

Another interesting holographic path to pursue would be the generalisation of our work to Kac-Moody algebras, similar to what was done in~\cite{Erdmenger:2020sup,Ghodrati:2019bzz} taking~\cite{Caputa:2018kdj} as a point of departure. In the bulk, this could be compared to volume calculations (similar to the comparison between the bulk results of \cite{Flory:2018akz} and the field-theory results of \cite{Belin:2018bpg}) in warped AdS$_{1+2}$ spaces along the lines of \cite{Ghodrati:2019bzz} or volumes calculations in the generalisation of Ba\~nados metrics under the boundary-conditions derived in \cite{Grumiller:2016pqb}.

Among other directions deserving further studies is certainly to shed light on the nature of the path-integral optimization~\cite{Caputa:2017urj,Caputa:2017yrh}. In this approach, one uses Euclidean path integrals to prepare states and views sources profiles -- in the best-explored case of CFT$_2$ this is the Weyl factor of the conformally-flat metric in which a CFT$_2$ lives -- as providing alternative ways of preparing certain states or operators. The optimal choice is the one that minimizes a desired cost functional. In~\cite{Camargo:2019isp} it was shown using~\cite{Milsted:2018yur} that the geometric functional adopted in~\cite{Caputa:2017urj,Caputa:2017yrh} can be viewed as an \emph{approximation} to an operator complexity cost function in which one counts insertions of the energy-momentum tensor operator\footnote{This discussion bears a striking similarity to our result~\eqref{ultralocal}.}. However, the meaning of the \emph{exact} cost function~\cite{Caputa:2017urj,Caputa:2017yrh} remained elusive. Since our present work establishes new results about cost functions associated with counting the energy-momentum tensor applications, the thread started with~\cite{Camargo:2019isp} certainly deserves another look. The first step to bridge in this direction is to view matrix elements of the circuit~\eqref{eq.circuit} with $Q$ given by~\eqref{eq.Qcft} as a Lorentzian path integral in a (1+1)-dimensional background geometry specified by~$f(\tau,\sigma)$.

A further interesting connection concerns the physics of Berry phases~\cite{Berry:1984jv} in the context of QFT complexity. To this end,~\cite{Erdmenger:2020sup} showed that, up to a boundary term, the circuit functional adopted in~\cite{Caputa:2018kdj} is directly related to a Berry phase\footnote{See also~\cite{Akal:2019hxa} in the context of complexity and Berry phases and~\cite{Oblak:2017ect,Oblak:2020jek} in the context of Berry phases and the Virasoro group. It is also interesting to note that Berry connection and the global conformal group appeared in~\cite{Czech:2019vih}, see also \cite{Chen:2020nlj} for a related development.}. It is well known that the structure of a real symmetric object (defining the quantum information metric \eqref{FSmetric}) added to an antisymmetric imaginary object (leading to the Berry phase) is very reminiscent of the quantum geometric tensor introduced in \cite{Berry:1989ai}. Given that in our calculations the~quantum information metric~\eqref{fsdefinition} is related to the two-point function of the energy-momentum tensor, it would be very interesting to identify the remaining part of the quantum geometric tensor and bring the discussion of Berry phases and state complexities in CFTs$_{1+1}$ into a unified conceptual framework.


\section*{Acknowledgements}

We would like to thank Volker Schomerus for being involved in the initial part of this collaboration, Pawe\l{} Caputa for discussions on \cite{Caputa:2018kdj} and Martin Ammon, Diptarka Das, Johanna Erdmenger,  Marius Gerbershagen, Romuald Janik, and Anna-Lena Weigel for conversations and correspondence. We are also grateful to Alexandre Belin, Minyong Guo, Ro Jefferson, Javier Mag\'{a}n, Ali Naseh, Onkar Parrikar, Blagoje Oblak, G\'{a}bor S\'{a}rosi and Claire Zukowski for useful comments on this draft. The Gravity, Quantum Fields and Information (GFQI) group at AEI is supported by the Alexander von Humboldt Foundation and the Federal Ministry for Education and Research through the Sofja Kovalevskaja Award. MF was supported by the Polish National Science Centre (NCN) grant 2017/24/C/ST2/00469. MF is also grateful to the organizers of the GQFI Workshop IV, where this work was presented for the first time, and the GFQI group at AEI for its hospitality.

\appendix

\section{$PSL(2,\mathbb{R})$ (gauge) symmetry and null circuits}
\label{sec::SL2R}

As pointed out in sections \ref{sec::deg} and \ref{sec::U1gauge}, the case $h=0$ is special as the degeneracy of the metric is enhanced by the appearance of the additional null-tangent vectors $v(\sigma)=\sin(\sigma+\delta)$, $0\leq\delta<2\pi$ at the point $f(\sigma)=\sigma$, and we expect this to lead to an enhanced  $PSL(2,\mathbb{R})$ 
(see \cite{Oblak:2016eij}) (gauge) symmetry of the geometric problem.

Following \cite{Oblak:2016eij}, we consider maps of the form
\begin{align}
    e^{i\sigma}\rightarrow e^{i G(\sigma)}=\frac{Ae^{i\sigma}+B}{\bar{B} e^{i\sigma}+\bar{A}},\ \  |A|^2-|B|^2=1.
    \label{SL2R0}
\end{align}
For infinitesimal transformations, we write $A=1+i \varsigma, B=\varrho$ with $\varsigma\in\mathbb{R}, \varrho\in\mathbb{C},\ \varsigma,|\varrho|\ll 1$. To first order, this yields
\begin{align}
G(\sigma)\equiv\sigma +g(\sigma)+\mO(\varsigma^2,\varrho^2)=\sigma +2 \varsigma +2\Im(\varrho) \cos(\sigma)-2\Re(\varrho)\sin(\sigma)+ \mO(\varsigma^2,\varrho^2),  
\label{SL2R}
\end{align}
and shows that the null-directions identified in section \ref{sec::deg} correspond to the generators of this group. Generalising the $U(1)$ transformations of section \ref{sec::U1gauge}, we hence consider transformations of the form
\begin{align}
f(\tau,\sigma)\rightarrow f(\tau,G(\tcs))=f(\tcs)+f'(\tcs)g(\tcs)+ \mO(\varsigma^2,\varrho^2)
\end{align} 
with $G(\tcs)$ given in \eqref{SL2R}, were we allow for a time dependence of the parameters $\varsigma(\tau),\varrho(\tau)$\footnote{Circuits of this form were also studied in  \cite{Erdmenger:2020sup} as solutions to their equations of motion.}. Under such an infinitesimal transformation, to first order the Lagrangian transforms as
  \begin{align}
 & \mathrm{L}_{sq}[f,f',\dot{f}]
	=\iint d\sigma 
	d\kappa  \Pi(\sigma-\kappa)  
	\frac{\dot{f}(\tcs)}{f'(\tcs)} \frac{\dot{f}(\tck)}{f'(\tck)}
		\label{SL2Raction}
\\
&\rightarrow \iint d\sigma 
	d\kappa  \Pi(\sigma-\kappa)  \Bigg(
	\frac{\dot{f}(\tcs)}{f'(\tcs)} \frac{\dot{f}(\tck)}{f'(\tck)}+\dot{g}(\tau ,\kappa )	\frac{\dot{f}(\tcs)}{f'(\tcs)}+\dot{g}(\tau ,\sigma ) \frac{\dot{f}(\tck)}{f'(\tck)}
	\nonumber
	\\
	&-\frac{\dot{f}(\tau ,\kappa ) \dot{f}(\tau ,\sigma ) g'(\tau ,\kappa )}{f'(\tau ,\kappa ) f'(\tau ,\sigma )}-\frac{\dot{f}(\tau ,\kappa ) \dot{f}(\tau ,\sigma ) g'(\tau ,\sigma )}{f'(\tau ,\kappa ) f'(\tau ,\sigma )}+\frac{\dot{f}'(\tau ,\kappa ) \dot{f}(\tau ,\sigma ) g(\tau ,\kappa )}{f'(\tau ,\kappa ) f'(\tau ,\sigma )}
	\nonumber
	\\
	&+\frac{\dot{f}(\tau ,\kappa ) \dot{f}'(\tau ,\sigma ) g(\tau ,\sigma )}{f'(\tau ,\kappa ) f'(\tau ,\sigma )}-\frac{f''(\tau ,\kappa ) \dot{f}(\tau ,\kappa ) \dot{f}(\tau ,\sigma ) g(\tau ,\kappa )}{f'(\tau ,\kappa )^2 f'(\tau ,\sigma )}-\frac{\dot{f}(\tau ,\kappa ) f''(\tau ,\sigma ) \dot{f}(\tau ,\sigma ) g(\tau ,\sigma )}{f'(\tau ,\kappa ) f'(\tau ,\sigma )^2}
	\Bigg).
\nonumber
\end{align}
Given the expression for $g(\sigma)$ from \eqref{SL2R}, the terms 
\begin{align}
\iint d\sigma 
d\kappa  \Pi(\sigma-\kappa)\left( \dot{g}(\tau ,\kappa )	\frac{\dot{f}(\tcs)}{f'(\tcs)}+\dot{g}(\tau ,\sigma ) \frac{\dot{f}(\tck)}{f'(\tck)}\right)
\end{align}
vanish under differential regularisation for $h=0$ \eqref{diffreg}.
	The remaining terms can be brought into the form
	  \begin{align}
 & \mathrm{L}_{sq}[f,f',\dot{f}]
	\rightarrow\iint d\sigma 
	d\kappa  \Pi(\sigma-\kappa)  
	\frac{\dot{f}(\tcs)}{f'(\tcs)} \frac{\dot{f}(\tck)}{f'(\tck)}
\\
&+\iint d\sigma 
	d\kappa  \frac{\dot{f}(\tcs)}{f'(\tcs)} \frac{\dot{f}(\tck)}{f'(\tck)} \Bigg(
	 \Pi '(\sigma -\kappa )\left(g(\tau ,\kappa )-g(\tau ,\sigma )\right)-2 \Pi (\sigma -\kappa )\left( g'(\tau ,\kappa )+ g'(\tau ,\sigma )\right)
	\Bigg).
\end{align}
Plugging in the concrete form for $g(\sigma)$ from \eqref{SL2R} (note that the time-dependence does not matter anymore) and 
\begin{align}
    \Pi(\sigma-\kappa)=\frac{c}{32\sin((\sigma-\kappa)/2)^4}
    \label{PIc}
\end{align}
we see that the expression in brackets in \eqref{SL2Raction} does indeed vanish identically, proving the invariance of the Lagrangian (for $h=0$) under the $PSL(2,\mathbb{R})$ gauge-symmetry. 
It is interesting to note here, however, that unlike the $U(1)$ invariance of section \ref{sec::U1gauge}, this symmetry is not a generic consequence of derivatives appearing in the differential regularisation of $\Pi$ (like equation \eqref{hatPI} where $\tilde{\Pi}$ is still arbitrary to a degree), but requires the specific form of \eqref{PIc}.

Following section \ref{sec::EA}, we will now discuss the issue of wave breaking for null-circuits in the $h=0$ case. Such a null-circuit can be written as \cite{Erdmenger:2020sup}
\begin{align}
f(\tcs)=2\arctan\left(\frac{\alpha(\tau)\tan(\sigma/2)+\beta(\tau)}{\gamma(\tau)\tan(\sigma/2)+\delta(\tau)}\right),\ \ \alpha(\tau)\delta(\tau)-\beta(\tau)\gamma(\tau)=1,
\end{align}
which is equivalent to \eqref{SL2R0}, but (more) manifestly real. Consider now the solution
\begin{align}
    \beta(\tau)=0,\  
    \alpha(\tau)=\gamma(\tau)+\delta(\tau),\ 
    \gamma(\tau)=\frac{1-\delta(\tau)^2}{\delta(\tau)},\ 
    \\
    \Rightarrow f(\tcs)=2 \cot ^{-1}\left(\delta(\tau) \left(\cot \left(\frac{\sigma}{2}\right)-1\right)+1\right).
\end{align}
It is easy to explicitly check that this circuit will indeed be null. Furthermore, we can calculate
\begin{align}
    f'(\tau,\pi/2)=\delta(\tau).
\end{align}
This means that with a convenient choice for $\delta(\tau)$, we can reach the boundary of the space of invertible maps ($\delta(\tau)\rightarrow+\infty$ or $\delta(\tau)\rightarrow 0$) in finite affine parameter $\tau$, and in fact the \textit{complexity distance} according to our metric between these boundary points and the identity map ($\delta(\tau)=1$) will be zero.
This is a general consequence of the topology of the $PSL(2,\mathbb{R})$ subgroup which, unlike the $U(1)$ subgroup, contains paths that (asymptotically) reach the boundary of the space of invertible maps.

\section{Fubini-Study complexity for $h/c\rightarrow\infty$}
\label{sec::effingprimes}

The case $h/c\rightarrow\infty$ may not be the most physical from a CFT point of view, but it does appear to have some peculiar properties that make it interesting from a purely geometric point of view. For example, as shown in section \ref{sec::sectional}, it is in this case that \textit{all} the sectional curvatures that we can calculate via the equivalent metric method are positive. Also, in this limit our equations may be related, via exchanging left for right-invariance, to the generalised CLM equation studied in \cite{Okamoto_2008,doi:10.1142/S1402925110000544,BAUER2016478}.

Taking the appropriate limit effectively amounts to substituting $c=0$ and we also scale away an unimportant overall factor of $h$ by setting it to unity. The result \eqref{sinsolution} yields
\begin{align}
f(\tcs)\approx &\sigma +\varepsilon\tau  \sin (\sigma )+\varepsilon^2\frac{5}{8} \left(\tau ^2-\tau \right)  \sin (2 \sigma )
\label{sinsolutionh}
\\
&+\varepsilon ^3 \left(\frac{-13 \tau ^3+30 \tau ^2-17 \tau}{48}\sin (\sigma )+\frac{29 \tau ^3-55 \tau ^2+26 \tau}{48} \sin (3 \sigma )\right)
\nonumber
\\
&+\varepsilon ^4 \Bigg(\frac{\left(-501 \tau ^4+1490 \tau ^3-1446 \tau ^2+457 \tau \right) \sin (2 \sigma )}{1152}
\nonumber
\\
&\ \ \ \ \ \ \ +\frac{\left(539 \tau ^4-1490 \tau ^3+1363 \tau ^2-412 \tau \right) \sin (4 \sigma )}{768} \Bigg)
\nonumber
\\
&+\varepsilon ^5\Bigg(\frac{\left(432 \tau^5-1925 \tau^4+2735 \tau^3-1270 \tau^2+28 \tau\right) \sin (\sigma )}{11520}
\nonumber
\\
&\ \ \ \ \ \ \ +\frac{\left(-24231 \tau^5+91250 \tau^4-126440 \tau^3+76240 \tau^2-16819 \tau\right) \sin (3 \sigma )}{34560}
\nonumber
\\
&\ \ \ \ \ \ \ +\frac{\left(3467 \tau^5-12550 \tau^4+16893 \tau^3-10004 \tau^2+2194 \tau\right) \sin (5 \sigma )}{3840}
\Bigg)
\nonumber
\\
&+...,
\nonumber
\end{align}
and the analogue of \eqref{epsilonseries} is
\begin{align}
\mathrm{L}_{sq}=2 \pi ^2 \varepsilon ^2+\frac{7 \pi ^2 \varepsilon ^4}{48}+\frac{3 \pi ^2 \varepsilon ^6}{80}+\frac{34693 \pi ^2 \varepsilon ^8}{2322432}+
\frac{31063061 \pi ^2 \varepsilon ^{10}}{4180377600}+\frac{560230064599 \pi ^2\varepsilon^{12}}{132434362368000}+...,
\label{epsilonseriesh}
\end{align}
where we have added a few additional orders which with some effort is doable in this simplified case. We are interested in whether the series in \eqref{epsilonseriesh} has a finite convergence radius $\varepsilon<1$ and describes an analytic function with a pole at $\varepsilon=1$, as we might conjecture based on the discussion at the end of section \ref{sec::approx}. It would thus be interesting to identify some regularity in the terms at each order $n$ in \eqref{epsilonseriesh} and deduce a candidate analytic function that has these terms as the initial terms of its Taylor-series, however, we have not been able to do so.

The attentive reader may have noticed that the first few numerators in equation \eqref{epsilonseriesh}, $2,7,3,34693$ and $31063061$, are all prime numbers. This trend is broken by the next term, as $560230064599=647\times 1433\times 604249$, however, we see that the prime-factors appearing here are still rather big. This is in stark contrast to the denominators, which have factorizations that tend to avoid large primes: $48=2^4\times3$, $80=2^4\times5$, $2322432=2^{12}\times3^4\times7$, $4180377600=2^{15}\times3^6\times5^2\times7$ and $132434362368000=2^{21}\times3^8\times5^3\times 7\times 11$. 
This feature is, with a little thought, not that surprising. Consider the expression
\begin{align}
B\equiv \frac{1}{N}\sum_{i=1}^N q_i
\label{B}
\end{align}
for some large integer $N$, and where the numbers $q_i\in \{q_i\} \subset \mathbb{Q}$ are rational numbers randomly picked from the finite set $\{q_i\}$ whose elements are \textit{approximately} evenly distributed in the interval $[0,1]$. A little bit of experimentation shows that the details of how the set $\{q_i\}$ is defined will not be important in general: 
The number $B$ will be similar to the terms in \eqref{epsilonseriesh} in the sense that the numerator will tend to have few large prime-factors while the denominator will tend to have many small prime-factors. This is a consequence of the summation of many terms, where the lowest common denominator has to be calculated each time two fractions are added. Hence, the denominator of $B$ will be a large number which is the product of many integers and hence has to be the product of many relatively small primes. However, as shown by the famous Erd\"{o}s-Kac theorem of probabilistic number theory, \textit{generic} large numbers are more likely to have few large prime factors than many small prime factors. Consequently, while both the numerator and the denominator of $B$ will be large numbers, the denominator which arises from products of smaller numbers will be special, while the numerator which arises from a sum of many terms will be generic in the sense of the Erd\"{o}s-Kac theorem.
Note that, qualitatively similar to \eqref{B}, the terms at each order in $\varepsilon$ in \eqref{epsilonseriesh} are the results of sums of increasingly many terms at increasing order. Not only does the function \eqref{sinsolutionh} itself contain increasingly many terms at increasing order in $\varepsilon$, the Taylor-series of $\frac{\dot{f}(\tcs)}{f'(\tcs)} \frac{\dot{f}(\tck)}{f'(\tck)}$ will likewise suffer a considerable proliferation of terms as the order in $\varepsilon$ is increased. Hence the appearance of large primes in \eqref{epsilonseriesh} is not surprising after all.  

\section{Equivalent metric}
\label{sec::inverse}


Investigating \eqref{EOMf2}, it would be helpful to find a function $\amalg(\alpha-\gamma)$ satisfying  
\begin{align}
\delta(\alpha-\beta)=\int_0^{2\pi}d\gamma \amalg(\alpha-\gamma)\Pi(\gamma-\beta).
\label{inverseproblem}
\end{align}
We could then left-multiply, i.e.~contract, \eqref{EOMf2} with an \textit{inverse metric} 
\begin{align}
\int d\kappa(\amalg(\rho-\kappa)f'(\tck)f'(\tau,\rho)\times...),
\end{align}
and would obtain an equation of the form
\begin{align}
\ddot{f}(\tau,\rho)\equiv\left(\text{other terms, at most first order in $d/d\tau$}\right),
\label{isolation}
\end{align}
which could be treated with standard numerical methods for initial value problems. It is then natural to approach~\eqref{inverseproblem} in Fourier space. For this, we use
\begin{align}
\Pi(\gamma-\beta)&
=\sum_{n\in\mathbb{Z}}\left(\frac{c}{24}(|n|^3-|n|)+|n|h\right)e^{-in(\gamma-\beta)}
\equiv\sum_{n\in\mathbb{Z}}\hat{\Pi}_n e^{-in(\gamma-\beta)}.
\label{2ptFourier}
\end{align}
Upon using Fourier representation for $\amalg$
\begin{align}
\amalg(\alpha-\gamma)&=\sum_{m\in\mathbb{Z}} \hat{\amalg}_m \, e^{im(\alpha-\gamma)},
\end{align}
we are led to
\begin{subequations}
\begin{align}
\delta(\alpha-\beta)=\frac{1}{2\pi}\sum_{n\in\mathbb{Z}} e^{in(\alpha-\beta)}&\equiv
\int_0^{2\pi}d\gamma \amalg(\alpha-\gamma)\,\Pi(\gamma-\beta)
\label{inverseproblem2}
\\
&=\sum_{n,m\in\mathbb{Z}}\int_0^{2\pi}d\gamma \,\hat{\amalg}_m\, \hat{\Pi}_n\, e^{i m \alpha} \,e^{-i\gamma(n+m)} \, e^{i n \beta}
\\
&=2\pi\sum_{m\in\mathbb{Z}}  \hat{\amalg}_m \,\hat{\Pi}_{-m} \, e^{i n(\alpha- \beta)},
\end{align}
\end{subequations}
which would require 
\begin{align}
\hat{\amalg}_m=\frac{1}{4\pi^2\,\hat{\Pi}_{m}}.
\label{amalg}
\end{align}
As can be seen from~\eqref{2ptFourier}, this expression would generically yield problems for the term $m=0$ and when $h = 0$ also for the terms at $m = \pm 1$.
So it seems that \eqref{inverseproblem} cannot be solved, which is just another consequence of the degeneracy of our metric discussed in sections \ref{sec::diffreg} and \ref{sec::U1gauge}. While the inverse metric of a degenerate metric cannot be defined, we will use what we refer to as the \textit{equivalent metric} method to define its substitute.\footnote{Alternatively, we could set the left-hand side of \eqref{inverseproblem} not equal to $\delta(\alpha-\beta)$, but instead $\delta'(\alpha-\beta)$ or in general a linear combination of sufficiently high derivatives of $\delta(\alpha-\beta)$. Then, the analogue of \eqref{inverseproblem2} could be solved for $\hat{\amalg}_m$ and the zeroes of $\hat{\Pi}_m$ would not cause a problem because of the additional factors of $n$ in the Fourier-expansion of  $\delta'(\alpha-\beta)$ leading to similar zeroes. Multiplying a "pseudo-inverse" $\amalg(\alpha-\gamma)$ defined this way to the equations of motion would formally allow us to isolate a combination of terms of the form $\ddot{f}(\tau,\rho)$, $\ddot{f}'(\tau,\rho)$, $\ddot{f}''(\tau,\rho)$ etc.~outside of the integral generalising \eqref{isolation}. As explained already in section \ref{sec::U1gauge}, in order to obtain $\ddot{f}(0,\sigma)$ for given initial conditions $f(0,\sigma),\dot{f}(0,\sigma)$ we would have to integrate this, potentially leading to an ambiguity. The occurrence of this ambiguity is equivalent to the problem coming from differential regularisation that we discussed at the end of section \ref{sec::VIP}, and fixing this ambiguity corresponds to fixing a gauge.}

To this end, consider first the problem of geodesic motion on a finite-dimensional degenerate metric, i.e.~a metric $g_{\mu\nu}$ with $\det(g)=0$ and hence an eigenvector $k^\mu$ with eigenvalue zero: $k^\mu g_{\mu\nu}=0$. We assume for simplicity that there is only one null eigenvector. The presence of~$k^{\mu}$ means we cannot straightforwardly define a non-trivial one-form $k_\nu$ by just lowering the index of $k^\mu$. We now fix a basis of tangent-vectors $\{Z_i^\mu,k^\mu\}$ such that 
\begin{align}
g_{\mu\nu}Z_i^\mu Z_j^\nu&=\delta_{ij} \quad \mathrm{and} \quad 
g_{\mu\nu}Z_l^\mu k^\nu=0\quad \forall l.
\end{align}
The vectors $\{Z_i^\mu\}$ are supposed to form a basis on a tangent-space for which a non-degenerate "equivalent metric" can be defined. This choice is not unique, as we could introduce shifts of the form $Z_i^\mu\rightarrow Z_i^\mu+k^\mu$. Fixing this ambiguity by making a choice of vectors $\{Z_i^\mu\}$ corresponds to fixing the ambiguities related to null-directions we have encountered in the earlier chapters, i.e.~gauge fixing. We can now define a non-zero one-form~$\bar{k}_\nu$ such that $Z_i^\mu \bar{k}_\mu=0\quad \forall i$. Note that in general $k^\mu\bar{k}_\mu\neq0$.

The geodesic equation for a (potentially) degenerate metric was given in \eqref{degenerategeodesics}. This cannot be brought into the usual form involving the Christoffel-symbols of the second kind $\Gamma^\gamma_{\alpha\beta}$ because the inverse $g^{\gamma\delta}$ of the degenerate metric does not exist. The expression 
\begin{align}
\Gamma_{\nu\alpha\beta}=\frac{1}{2}\left(g_{\alpha\nu,\beta}+g_{\beta\nu,\alpha}-g_{\alpha\beta,\nu}\right)
\label{christoffel}
\end{align}
is of course readily recognised as the Christoffel-symbols of the first kind or Koszul object~\cite{Stoica:2013wx}. 
We will now make the assumption that we are interested exclusively in geodesics $X^\mu(\tau)$ that move entirely within the selected tangent-space spanned by $\{Z_i^\mu\}$, i.e.~$\dot{X}^\mu \bar{k}_\mu = \ddot{X}^\mu\bar{k}_\mu = 0$. Then we can show that the equations of motion \eqref{degenerategeodesics} are still satisfied by such geodesic curves even if we shift the metric by
\begin{align}
g_{\mu\nu}\rightarrow g^{eq}_{\mu\nu}=g_{\mu\nu}+\lambda \, \bar{k}_\mu\bar{k}_\nu,
\label{effective metric}
\end{align}
where $\lambda$ is an arbitrary function. Hence for the geodesics of interest,
the metric $g_{\mu\nu}$ is equivalent to the metric $g^{eq}_{\mu\nu}$ which we aptly call the equivalent metric. The benefit now is that this metric will in general be invertible, allowing us to compute the Christoffel symbols of the second kind as well as the Riemann and Ricci-tensors and the Ricci scalar. The sectional curvatures calculated from this Riemann tensor will then accurately describe the geometric behaviour of the geodesic curves of interest. 

Coming back to our infinite dimensional case with $h\neq0$, in order to define an equivalent metric we have to find a function $\bar{k}(\sigma)$ such that 
\begin{align}
\int d\sigma \,\bar{k}(\sigma)\,\dot{f}(\tcs) =  \int d\sigma\, \bar{k}(\sigma)\,\ddot{f}(\tcs) = 0
\end{align}
for those circuits $f(\tcs)$ that satisfy whatever gauge-condition we have chosen. If we require for example
\begin{equation}
\label{eq.gaugecondfeq0}
f(\tau,0)=0\quad \forall\tau,    
\end{equation} 
as suggested at the end of section \ref{sec::U1gauge}, we would be led to $\bar{k}(\sigma)\sim\delta(\sigma)$ and the equivalent metric would be obtained by the shift
\begin{align}
\frac{\Pi(\sigma-\kappa)}{f'(\tcs)f'(\tck)}\rightarrow\frac{\Pi(\sigma-\kappa)}{f'(\tcs)f'(\tck)}+\lambda\frac{\delta(\sigma)\delta(\kappa)}{f'(\tcs)f'(\tck)}.
\end{align}

Instead of this, let us make a slightly more complicated ansatz and demand that we are only interested in circuits $f(\tcs)$ that the following condition
\begin{align}
\int d\sigma \frac{\dot{f}(\tcs)}{f'(\tcs)} =0\ \forall \tau
\label{condition}
\end{align}
instead of~\eqref{eq.gaugecondfeq0}. It is an easy task to show that for such circuits, the two-point function in \eqref{EOMf2} can be replaced by the equivalent two-point function
\begin{align}
\Pi(\sigma-\kappa)\rightarrow\Pi^{eq}(\sigma-\kappa)=\Pi(\sigma-\kappa)+\lambda,
\label{eqmetriclambda}
\end{align}
where $\lambda$ is a constant, and those solutions to the equations of motion that satisfy \eqref{condition} will not cease to be solutions to the equations of motion for $\lambda\neq 0$. This shift of the two-point function by a constant is exactly what is needed to make equation \eqref{amalg} well defined, as $\hat{\Pi}_0\neq0$ for $\lambda\neq0$, so in a sense this is the easiest equivalent metric we can study in our problem, which we will do in section \ref{sec::sectional}. 

We close this section by investigating how for a given circuit $f(\tcs)$ the "gauge" \eqref{condition} can be enforced. Suppose that initially
\begin{align}
\int d\sigma \frac{\dot{f}(\tcs)}{f'(\tcs)} =\beta(\tau). 
\end{align}
Now, a $U(1)$ gauge transformation defined in section~\ref{sec::U1gauge} leads to
\begin{subequations}
\begin{align}
&f(\tcs)\rightarrow f(\tau,\sigma+\alpha(\tau)),
\\
&f'(\tcs)\rightarrow f'(\tau,\sigma+\alpha(\tau)),
\\
&\dot{f}(\tcs)\rightarrow \partial_\tau f(\tau,\sigma+\alpha(\tau))= \dot{f}(\tau,\sigma+\alpha(\tau))+f'(\tau,\sigma+\alpha(\tau))\dot{\alpha}(\tau).
\end{align}
\end{subequations}
Note that in the last line, we use the dot in $\dot{f}$ to denote the derivative with respect the the first argument of the function $f$. 
 Consequently 
\begin{align}
\int d\sigma \frac{\dot{f}(\tcs)}{f'(\tcs)} \rightarrow \int d\sigma \left(\frac{\dot{f}(\tau,\sigma+\alpha(\tau))}{f'(\tau,\sigma+\alpha(\tau))} +\dot{\alpha}(\tau)\right)=\beta(\tau)+2\pi \dot{\alpha}(\tau)\overset{!}{=}0
\end{align}
Clearly, this can be set to zero solving the last equation as a differential equation for $\alpha(\tau)$. However, since this is a first order equation, the condition of the form $\alpha(\tau_0)=0$ can only be accommodated in general at either the beginning or the end of the circuit, but not both.

\bibliography{main}{}
\bibliographystyle{JHEP}

\end{document}